# Hierarchical On-Surface Synthesis of Deterministic Graphene Nanoribbon Heterojunctions

Christopher Bronner*,1,#, Rebecca A. Durr*,2, Daniel J. Rizzo1, Yea-Lee Lee1,3, Tomas Marangoni2, Alin Miksi Kalayjian2, Henry Rodriguez1, William Zhao1, Steven G. Louie1,4, Felix R. Fischer2,4,5,&, Michael F. Crommie1,4,5,#

[1]Department of Physics, University of California, Berkeley, CA 94720, USA. [2]Department of Chemistry, University of California, Berkeley, CA 94720, USA. [3]Department of Physics, Pohang University of Science and Technology, Pohang, Kyungbuk, 37673, Korea. [4]Materials Sciences Division, Lawrence Berkeley National Laboratory, Berkeley, CA 94720, USA. [5]Kavli Energy NanoSciences Institute at the University of California Berkeley and the Lawrence Berkeley National Laboratory, Berkeley, California 94720, USA.
* These authors contributed equally to this work.
# Corresponding authors

**Bottom-up graphene nanoribbon (GNR) heterojunctions are nanoscale strips of graphene whose electronic structure abruptly changes across a covalently bonded interface. Their rational design offers opportunities for profound technological advancements enabled by their extraordinary structural and electronic properties. Thus far the most critical aspect of their synthesis, the control over sequence and position of heterojunctions along the length of a ribbon, has been plagued by randomness in monomer sequences emerging from step-growth copolymerization of distinct monomers. All bottom-up GNR heterojunction structures created so far have exhibited random sequences of heterojunctions and, while useful for fundamental scientific studies, are difficult to incorporate into functional nanodevices as a result. Here we describe a new hierarchical fabrication strategy that allows *deterministic growth* of bottom-up GNRs that preferentially exhibit a single heterojunction interface rather than a random statistical sequence of junctions along the ribbon. Such heterojunctions provide a viable new platform that could be directly used in functional GNR-based device applications at the molecular scale. Our hierarchical GNR fabrication strategy is based on differences in the dissociation energies of C–Br and C–I bonds that allow control over the growth sequence of the block-copolymers from which GNRs are formed, and consequently yields a significantly higher proportion of single-junction GNR heterostructures. Scanning tunnelling spectroscopy and density functional theory calculations confirm that hierarchically-grown heterojunctions between chevron GNR (cGNR) and binaphthyl-cGNR segments exhibit straddling Type I band alignment in structures that are only one atomic layer thick and 3 nm in width.**

Functional GNRs are attractive candidates for high-speed digital nanodevices because they develop sizable bandgaps (e.g., 1 – 3 eV) as their widths become small (e.g., 1 – 3 nm).[1,2,3,4,5] For example, GNR-based heterojunctions could be employed in devices such as molecular-scale tunnelling field effect transistors and resonant tunnelling diodes.[6,7,8,9] The extreme sensitivity of GNR electrical properties to minute structural variations,[1,2,3,4,5,10] however, requires that practical GNR heterojunctions must have feature sizes that are well-controlled at the atomic scale. While this represents an insurmountable challenge for current top-down fabrication techniques,[11,12,13,14] it is actually routine practice using new molecular assembly-based bottom-up techniques that involve on-surface polymerization of molecular precursors followed by cyclodehydrogenation.[15,16,17,18,19,20,21,22] Atomically-precise bottom-up GNR heterojunctions have been synthesized previously in this way by combining molecular precursors that have different heteroatom doping patterns[23] (leading to dopant-induced heterojunctions) or different widths[24] (leading to width-based heterojunctions). GNR heterojunctions have also been fabricated from a single molecular precursor designed with sacrificial ligands that can be removed or chemically altered after growth to create abrupt variations in bandgap profile along the GNR axis.[25,26] The number and placement of all such heterojunctions, however, has so far been random due to the stochastic nature of thermally-driven molecular step growth polymerization, a situation that is problematic for the design, reproducibility, and ultimately implementation of functional GNR heterojunction nanodevices.

Hierarchical growth provides a potential solution to this problem as it provides an additional level of control to the bottom-up synthesis approach.[27,28] This arises from the fact that a careful selection of molecular building blocks can lead to a sequential activation of the growth process at different temperatures. Thermally-driven GNR self assembly is still a random process even under hierarchical growth conditions, but *different molecules* are induced to polymerize at *different temperatures*, thus providing an added element of control. For example, if a first GNR assembles (i.e., polymerizes) at temperature $T_1$ and a second GNR assembles at temperature $T_2 > T_1$, then by ramping the temperature it should be possible to grow a heterojunction between both GNRs that has only a single interface rather than the stochastic interfaces expected from a mixture of different precursors that polymerize at the same temperature. There are three elements needed to successfully achieve this novel growth process: (i) precursor molecules for the first GNR that can be activated at $T_1$, (ii) precursor molecules for the second GNR that polymerize at $T_2 > T_1$, and (iii) a linker molecule that facilitates growth of the second GNR off the end of the first GNR. Our strategy for accomplishing this hierarchical growth relies on functionalizing the first GNR precursors with iodine and the second GNR precursors with bromine. Because the C–I bond is weaker than the C–Br bond,[29,30,31,32,33] this ensures that $T_2 > T_1$ (due to the fact that GNR polymerization does not occur until thermally-driven dehalogenation causes precursors to become reactive radicals). Our linker elements are precursors of the second GNR that are functionalized on one side with iodine and on the other side with bromine groups. A related strategy has been developed previously to facilitate the growth of 2D polymer networks, but with no heterojunction functionality.[27,28]

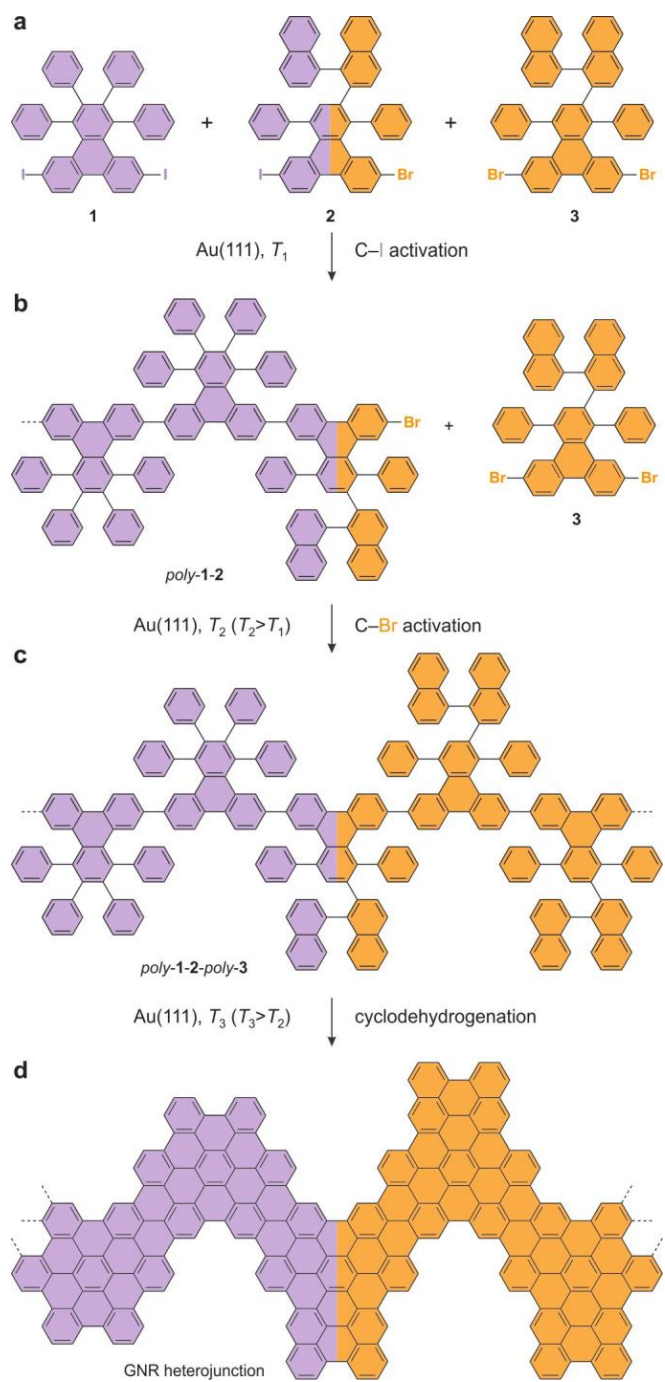

**Figure 1 | Schematic representation of the hierarchical on-surface synthesis of GNR heterojunctions. a**, Molecular precursors **1**, **2,** and **3**. **b**, Selective activation of the C–I bonds at $T_1$ leads to *poly*-**1**-**2** terminated by the bifunctional linker **2**. C–Br bonds in **2** and **3** are not activated at $T_1$. **c**, Selective activation of the C–Br bonds at $T_2$ results in a block-copolymer consisting of *poly*-**1** and *poly*-**3** segments fused by the linker **2**. **d**, Cyclodehydrogenation at $T_3$ yields a GNR heterojunction between fully cyclized cGNR and binaph-cGNR segments.

**Hierarchical On-Surface Synthesis of Controlled GNR heterojunctions**

For the hierarchical growth demonstrated here the first GNR was chosen to be the well-known chevron GNR[15,34,35,23,36] (cGNR) whereas the second GNR is a new chevron structure (binaphthyl-cGNR (binaph-cGNR)) that was designed to have a wider spatial profile specifically for this study. The chosen linker element possesses the same structure as binaph-cGNR. Fig. 1 shows the reaction of the iodinated precursor **1** that gives rise to the cGNR, the dual-functionalized linker precursor **2**, and the brominated precursor **3** that gives rise to the binaph-cGNR. At the lower polymerization temperature ($T_1$) only the C–I bond in **2** will be activated and the linker molecule will effectively terminate the growth of homopolymers of **1** (Fig. 1b). At $T_2$ the C–Br bond in **2** can be cleaved and serves as a seed for the polymerization of **3** (Fig. 1c). Precursor **3** will induce polymerization of the wider binaph-cGNR segment at temperatures higher than those required for the polymerization of the iodinated precursors **1** and **2**. GNR heterojunctions arising after cyclodehydrogenation of the block-copolymers arising from this process at $T_3$ are expected to show type I band alignment due to the smaller bandgap that arises from the wider binaphthyl-GNR segment.

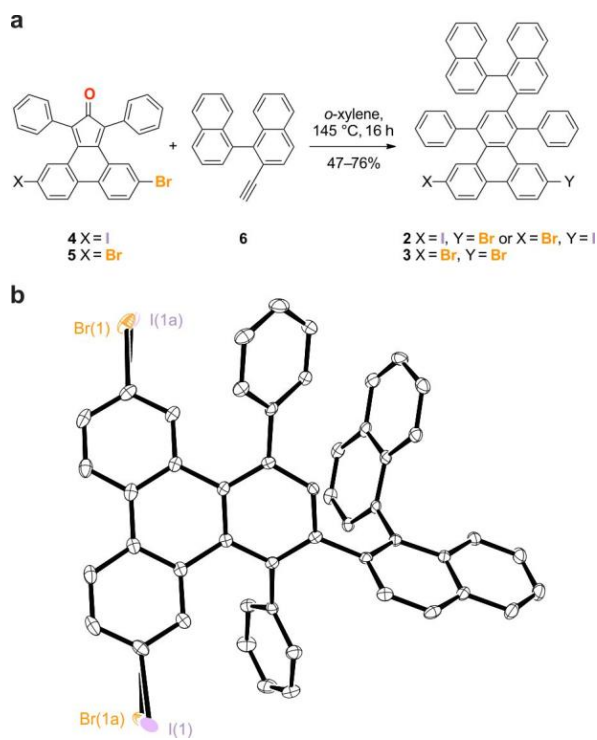

**Figure 2 | Synthesis of bifunctional linker 2 and binaph-cGNR precursor 3. a**, Synthesis of linker **2** and binaph-cGNR precursor **3**. **b**, ORTEP representation of the X-ray crystal structure of **2**. Thermal ellipsoids are drawn at the 50% probability level. Colour coding: C (grey), Br (orange), I (purple). There is a 50% compositional disorder between the Br(1) and I(1) sites. Hydrogen atoms are omitted for clarity.

The synthesis of **1** has been reported elsewhere.[37,36] Precursor **2** was obtained through monoiodination of 9,10-phenanthrenequinone followed by bromination to yield 2-bromo-7-iodophenanthrene-9,10-dione in 13% yield. Knoevenagel condensation with 1,3-diphenyl acetone followed by a Diels-Alder reaction with 2-ethynyl-1,1'-binaphthalene (**6**) afforded **2** in 47% yield (Fig. 2a). Precursor **3** was prepared from 2,7-dibromophenanthrene-9,10-dione in two steps in 76% yield. $^1$H NMR of analytically pure samples of **2** and **3** at 24 °C reveals a complex spectrum attributed to the slow interconversion of rotational isomers around the binaphthyl group. Variable temperature NMR of both **2** and **3** in 1,1,2,2-tetrachloroethane-$d_2$ at 110 °C resolves the spectroscopic signals (Supplementary Fig. S6 and S7). Pale yellow crystals of **2** suitable for X-ray diffraction were grown from saturated CHCl$_3$/MeOH solutions. In the crystal structure **2** exhibits a 50% compositional disorder between the Br(1) and I(1) sites (Fig. 2b), revealing a 1:1 mixture of constitutional isomers of **2** based on the connectivity of the binaphthyl group to the triphenylene core at positions C(1) and C(2).

The hierarchical on-surface growth protocol was implemented in four steps. Step I: molecular precursors **1**–**3** were sequentially deposited onto a clean Au(111) substrate (**1** and **3** were deposited in excess relative to **2**). Step II: the surface temperature was raised to $T_1$ in order to induce the homolytic cleavage of the C–I bonds in **1** and **2**, thus leading to the formation of linear chains of *poly*-**1** terminated by the linker molecule **2** (*poly*-**1**-**2** in Fig. 1b). Step III: the surface temperature was raised further to $T_2$ in order to activate the C–Br bonds in **2** and **3**, thus extending the polymer chains from the ends of *poly*-**1**-**2**. The monomer sequence in the resulting block-copolymer (Fig. 1c) determines the segmentation of the GNR heterostructure. Step IV: the surface temperature was raised to $T_3$ to induce the cyclodehydrogenation reaction leading to fully extended GNRs featuring a single in-line heterojunction (Fig. 1d).

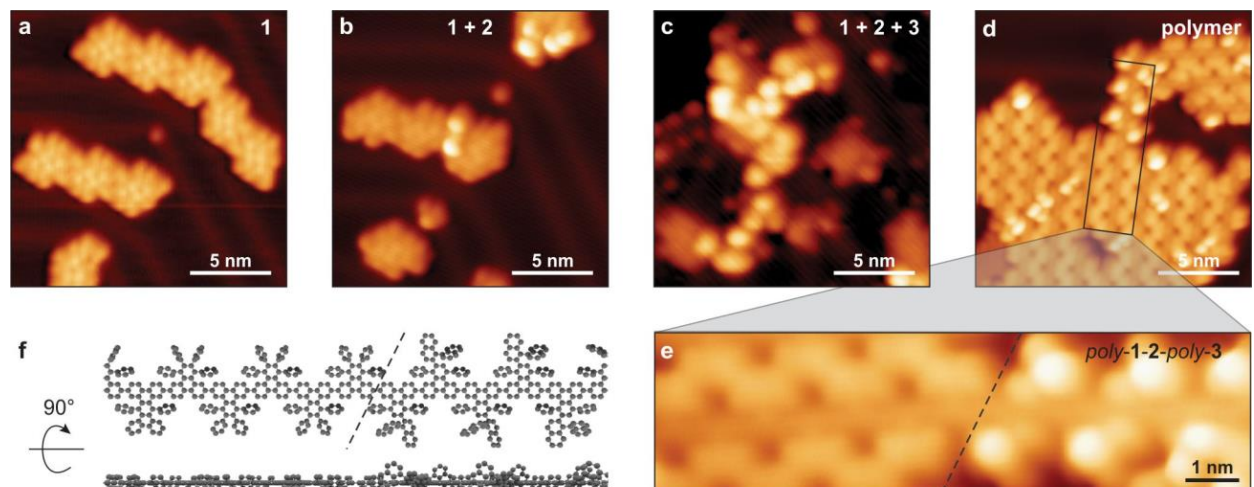

**Figure 3 | Hierarchical on-surface copolymerization. a**, STM topograph of **1** adsorbed on Au(111) ($V$ = 1.5 V, $I$ = 20 pA). **b**, **1** and **2** co-adsorbed on Au(111) ($V$ = 1.5 V, $I$ = 20 pA). Taller protrusions correspond to **2**. **c**, **1**, **2**, and **3** co-adsorbed on Au(111) ($V$ = 1.5 V, $I$ = 20 pA). **d**, Island of copolymers on Au(111) after annealing to 200 °C ($V$ = 1.0 V, $I$ = 20 pA). **e**, Zoom-in of copolymer outlined in **d** ($V$ = 1.0 V, $I$ = 20 pA). **f**, Top-down and side-on view of a molecular model for the copolymer depicted in **e**.

Scanning tunnelling microscopy (STM) imaging was used to follow the experimental implementation of this growth protocol. Figs. 3a-c depict the results of Step I. Fig. 3a shows the Au(111) surface after deposition of **1** while Fig. 3b shows the surface after additional deposition of small amounts of **2**. The molecular precursors cluster along the Au(111) herringbone reconstruction. The apparent height of **2** (4.9 Å) is significantly larger than the height of **1** (2.6 Å) due to the nonplanar arrangement of the binaphthyl group in **2**. This unique structural feature allows a clear distinction between monomers of **1** and **2** on the surface. Fig. 3c shows an STM image of the surface following the additional deposition of **3**.

Steps II and III were performed by gradually increasing the temperature of the surface to 200 °C at a rate of 2 K min$^{-1}$. Fig. 3d shows an STM image of the resulting polymers self-assembled into ordered islands, similar to the polymer stage for pure cGNRs.[15,36] The polymers exhibit segments with different apparent heights, thus allowing taller binaphthyl-containing segments to be distinguished from other chevron polymer segments (see close-up in Fig. 3e). A structural model in Fig. 3f illustrates the binaphthyl groups protruding from the molecular plane.

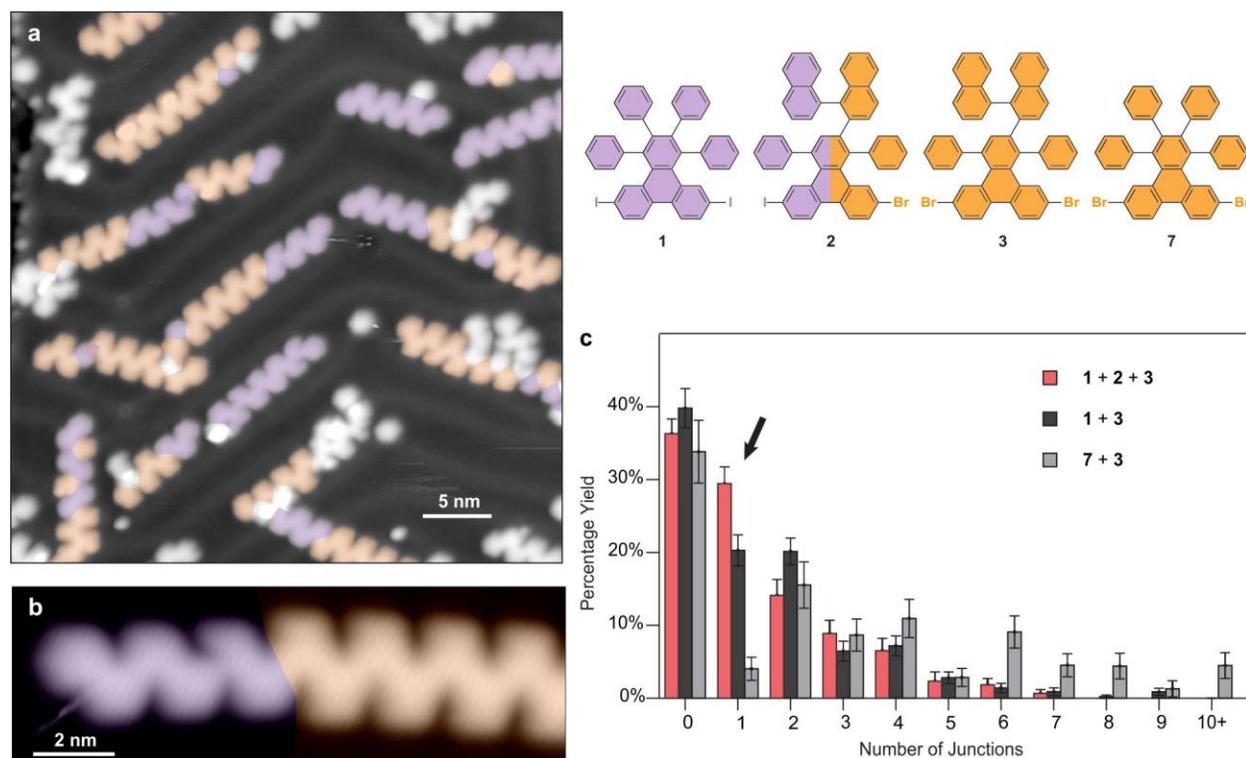

**Figure 4 | Hierarchically grown GNR heterojunctions. a**, STM image of cGNR/binaph-cGNR heterojunctions ($V$ = 0.3 V, $I$ = 20 pA). cGNR (binaph-cGNR) segments are highlighted in purple (orange). **b**, Magnified STM image of a GNR heterojunction ($V$ = 0.3 V, $I$ = 20 pA). **c**, Relative occurrence of GNRs containing different numbers of heterojunctions upon synthesis from precursors **1–3** ("full hierarchical protocol", red), **1** and **3** ("partial hierarchical protocol", dark grey), and **7** and **3** ("random protocol", light grey). The black arrow emphasizes the increase in single-junction GNR heterostructures arising from hierarchical growth.

Step IV was accomplished by ramping the sample temperature to 340 °C. As seen in Fig. 4a, this results in isolated GNRs comprised of fully cyclized cGNR and binaph-cGNR segments joined in heterojunction structures. STM topography of the segmented GNRs shows a preferred alignment with the herringbone reconstruction and a uniform apparent height of 2.2 Å, lower than the polymers and consistent with previous GNR measurements.[15,16] While the median length of the GNRs was 8 nm, some GNRs exhibited lengths exceeding 20 nm (see Supplementary Fig. S3). cGNR segments and binaph-cGNR segments can clearly be distinguished based on differences in width and shape and are highlighted in purple (from monomer **1**) and orange (from linker **2** and monomer **3**) colours in Fig. 4a (see Supplementary Fig. S4 for unprocessed image). Some GNRs can be found on the surface that are comprised of homopolymers of pure chevron or pure binaphthyl building blocks, but most contain a heterojunction.

In order to assess quantitatively whether the hierarchical growth strategy introduced here provides additional control over heterojunction formation compared to random heterojunction synthesis (as performed previously[23,24]), we conducted two control experiments and compared the heterojunction statistics. In the first control experiment (the "partial hierarchical protocol") we synthesized GNR heterostructures by co-depositing only **1** and **3** (thus omitting the linker molecule **2**) and then following the same annealing protocol as before. The purpose here was to test the advantage of using the bifunctional linker molecule **2** to cap *poly*-**1** in order to promote single-heterojunction formation. In the second control experiment (the "random protocol") we co-deposited the brominated binaphthyl precursor **3** with a conventional brominated cGNR precursor **7** (see Fig. 4c), and then followed the same annealing protocol. This second control experiment is essentially the same technique used previously to create random heterojunctions from two different precursors.[23,24]

Fig. 4c shows a histogram depicting the relative abundance of GNRs containing a given number of heterojunctions for all three growth protocols (i.e., the full hierarchical growth protocol as well as the two control experiments). A comparison of the three procedures confirms that the full hierarchical growth protocol does indeed result in a significant increase of single-junction GNRs. This effect is largest compared to the random protocol where the relative number of single-junction GNRs is increased by a factor of seven. The full hierarchical growth protocol also results in a 45% increase in single-junction GNRs compared to the partial hierarchical protocol. Overall, hierarchical growth is seen to provide significantly better control over GNR heterojunction synthesis.

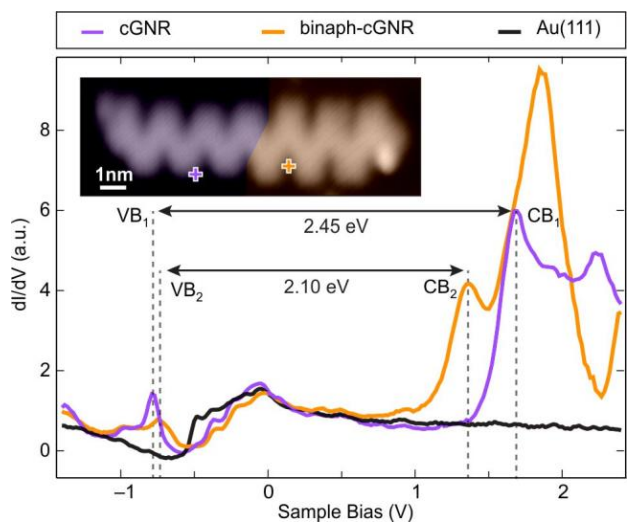

**Figure 5 | Spatially inhomogeneous bandgap of cGNR/binaph-cGNR heterojunction.** d*I*/d*V* spectra recorded above the cGNR (purple) and binaph-cGNR (orange) segments of a single-junction GNR compared to the bare

Au(111) surface. VB$_{1,2}$ and CB$_{1,2}$ denote the valence and conduction band edges of the cGNR and binaph-cGNR segments, respectively. Inset: STM image showing spectroscopy locations (STS set point: $V$ = 0.3 V, $I$ = 20 pA).

**Electronic Characterization of GNR Heterojunctions**

The electronic structure of GNR heterojunctions prepared following the hierarchical growth protocol was characterized using a combination of d$I$/d$V$ spectroscopy and density functional theory (DFT) simulations. Fig. 5 shows the results of d$I$/d$V$ spectroscopy performed on a narrow cGNR heterojunction segment and a wide binaph-cGNR segment within the single-junction GNR shown in the inset. Both segments exhibit peaks indicative of the energies of the valence band (VB) edge and conduction band (CB) edge of the respective segments. In the cGNR (binaph-cGNR) segment, the VB edge lies at $E_{VB1}$ = −0.78 ± 0.03 eV ($E_{VB2}$ = −0.74 ± 0.04 eV) while the CB edge lies at $E_{CB1}$ = 1.67 ± 0.03 eV ($E_{CB2}$ = 1.36 ± 0.02 eV) with respect to the Fermi level. The resulting bandgap of 2.45 ± 0.05 eV in the cGNR segment agrees well with previous scanning tunnelling spectroscopy (STS) experiments[26] and reasonably well with other reported values in the literature ranging from 2.0 eV to 3.1 eV.[23,35,34] The binaph-cGNR segment, which has not been reported before, features a smaller bandgap of 2.10 ± 0.05 eV. The reduction of the band gap is consistent with the extension of the conjugated π-system in the binaph-cGNR segment. The straddling band alignment of the two GNR segments defines the heterojunction as Type I.

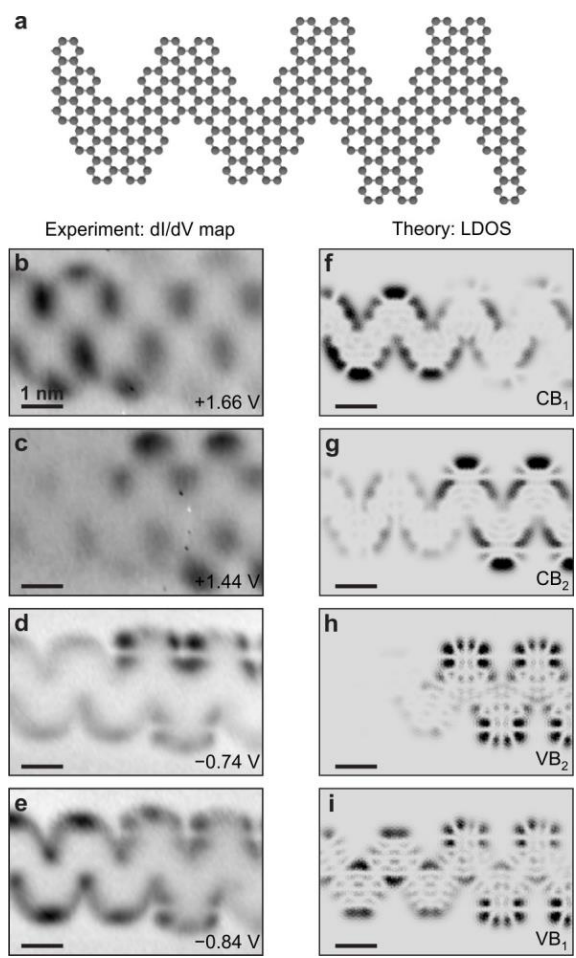

**Figure 6 | Spatial distribution of heterojunction band edge states. a**, Molecular model of the GNR heterojunction investigated in **b-i**. **b-e,** Constant-current d*I*/d*V* maps of VB and CB edge states for the two heterojunction segments. **f-i**, Simulated local density of states associated with VB and CB edge states for the two heterojunction segments calculated using DFT. Black (grey) represents high (low) intensity.

A characteristic feature of heterojunctions is that band edge wavefunctions tend to localize on one side of the heterojunction interface.[23] We explored this behaviour in hierarchically grown cGNR/binaph-cGNR heterojunctions using d*I*/d*V* mapping. Fig. 6 shows the wavefunction distribution (i.e., local density of states (LDOS)) across the heterojunction interface for states at the four band edges of the two GNR segments. Spatial localization is clearly observed at the band edges of the binaph-cGNR segment ($CB_2$, $VB_2$) where the wavefunction appears more intense on the binaph-cGNR side of the interface (Figs. 6c,d). Some localization is also seen at the cGNR conduction band edge ($CB_1$) where the wavefunction is more intense on the cGNR side of the interface (Fig. 6b). The wavefunction at the valence band edge of the cGNR ($VB_1$), however, does not show significant localization across the interface (Fig. 6e).

In order to test that our experimental cGNR/binaph-cGNR heterojunction behaviour is consistent with the expected electronic properties of a GNR heterojunction, we performed *ab initio* simulations of the heterojunction electronic structure using DFT. Supplementary Fig. S5 shows the unit cell used in the calculation. Our simulation confirms the reduced energy gap of the binaph-cGNR segment compared to the cGNR segment and reproduces the Type I heterojunction band alignment seen experimentally (Supplementary Fig. S5). Theoretical simulations of the spatial distribution of the heterojunction LDOS (Figs. 6f-i) confirm the wavefunction localization observed experimentally. Theoretical LDOS maps obtained at the band edge energies of the binaph-cGNR segment ($CB_2$, $VB_2$) show noticeable localization on that side of the heterojunction interface, as well as a distinct nodal structure that resembles the experimental LDOS distribution (Figs. 6c,d). Like the experiment, the theoretical LDOS distribution at the cGNR conduction band energy ($CB_1$) shows some localization on the cGNR side (Fig. 6f) while the LDOS at the cGNR valence band energy ($VB_1$) shows no discernible localization (Fig. 6i). The reduced wavefunction localization at the cGNR band edges arises from the fact that they are degenerate with states in the binaph-cGNR segment due to the Type I heterojunction band alignment. The band edge states on the binaph-cGNR side ($CB_2$, $VB_2$), by contrast, lie in the cGNR gap, and are thus more strongly confined.

**Conclusion**

We have demonstrated hierarchical on-surface synthesis of GNR heterojunctions from molecular precursors engineered to yield a predetermined growth sequence. This was accomplished by taking advantage of the subtle differences in the bond dissociation energies of C–I bonds compared to C–Br bonds to separate polymerization temperatures for different precursors. We observe that the use of a bifunctional linker molecule (i.e., one that includes both C–I *and* C–Br bonds) leads to single-heterojunction yields that are dramatically improved when compared to more standard uniform precursor functionalization (i.e., the use of exclusively brominated precursors), and also significantly better than a partial hierarchical growth protocol that forgoes the linker. STS measurements on hierarchically grown cGNR/binaph-cGNR heterojunctions reveal a Type I band alignment with strong wavefunction localization for the bands closest to the Fermi energy, consistent with *ab initio* simulations. The improved GNR heterojunction structural control demonstrated here for hierarchical growth techniques paves the way toward integrating atomically-precise GNR heterostructures into new nanoelectronic devices.

**Methods**

**Precursor Synthesis.** Full details regarding the synthesis and characterization of all precursor materials are given in the Supplementary Information.

**STM measurements.** The on-surface reactions were conducted on a clean Au(111) single crystal which was prepared by cycles of Ar$^+$ sputtering and annealing. All precursor molecules were evaporated from home-made Knudsen cells. The Knudsen cell sublimation temperatures of the precursors were 170 °C (**1**), 200 °C (**2**), 165 °C (**3**) and 160 °C (**7**). The monomers were deposited onto the substrate while holding it at $T < -50$ °C (the sample was taken directly from the cryogenic STM stage just prior to evaporation). STM imaging was performed in constant-current mode using a home-built STM at a temperature of $T = 13$ K. Differential conductance (d$I$/d$V$) measurements were recorded using a lock-in amplifier with modulation frequency of 566 Hz and modulation amplitude $V_{rms} = 10-25$ mV. d$I$/d$V$ point spectra were recorded under open feedback loop conditions. d$I$/d$V$ maps were collected under constant-current conditions.

**Calculations.** Theoretical simulations of freestanding GNR heterojunctions were performed using DFT within the local density approximation (LDA) as implemented in the Quantum Espresso package.[38] A supercell with sufficient vacuum space (> 10 Å) was used to avoid spurious interaction between periodic replicas. We used norm-conserving pseudopotentials with a planewave energy cut-off of 60 Ry. The heterojunction structure was fully relaxed until the force on each atom was smaller than 0.025 eV/Å. All edges were saturated with hydrogen atoms. A Gaussian broadening of 0.08 eV was used in the LDOS calculations.

**Acknowledgments**

Research supported by the Office of Naval Research MURI Program N00014-16-1-2921 (precursor design, surface growth, STM imaging, band structure), by the U.S. Department of Energy (DOE), Office of Science, Basic Energy Sciences (BES) under the Nanomachine Program award no. DE-AC02-05CH11231 (STM spectroscopy and simulation) and award no. DE-SC0010409 (precursor synthesis and characterization), by the Center for Energy Efficient Electronics Science NSF Award 0939514 (heterojunction modelling), and by the National Science Foundation under grant DMR-1508412 (development of theory formalism). Computational resources have been provided by the DOE at Lawrence Berkeley National Laboratory's NERSC facility and by the NSF through XSEDE resources at NICS. C.B. acknowledges support through the Fellowship Program of the German National Academy of Sciences Leopoldina under grant no. LPDS 2014-09.

**Author contributions**

C.B., R.A.D., M.F.C., F.R.F. conceived the experiments, R.A.D., T.M., A.M.K. designed, synthesized, and characterized the molecular precursors, C.B., D.J.R., H.R. and W.Z. performed the on-surface synthesis and STM characterization, Y.-L. L. and S.G.L. performed and interpreted the DFT calculations. C.B., R.A.D., D.J.R., Y.-L. L., S.G.L., F.R.F. and M.F.C. wrote the manuscript. All authors contributed to the scientific discussion.

**Additional information**

**Competing financial interests**

The authors declare no competing financial interests.

# Supplementary Information

# Hierarchical On-Surface Synthesis of Deterministic Graphene Nanoribbon Heterojunctions

**Christopher Bronner*,1,#, Rebecca A. Durr*,2, Daniel J. Rizzo1, Yea-Lee Lee1,3, Tomas Marangoni2, Alin Miksi Kalayjian2, Henry Rodriguez1, William Zhao1, Steven G. Louie1,4, Felix R. Fischer2,4,5,#, Michael F. Crommie1,4,5,#**

[1]Department of Physics, University of California, Berkeley, CA 94720, USA. [2]Department of Chemistry, University of California, Berkeley, CA 94720, USA. [3]Department of Physics, Pohang University of Science and Technology, Pohang, Kyungbuk, 37673, Korea. [4]Materials Sciences Division, Lawrence Berkeley National Laboratory, Berkeley, CA 94720, USA. [5]Kavli Energy NanoSciences Institute at the University of California Berkeley and the Lawrence Berkeley National Laboratory, Berkeley, California 94720, USA.
\* These authors contributed equally to this work.
\# Corresponding authors

# 1. Control Experiments Demonstrating the Effect of Hierarchical Growth

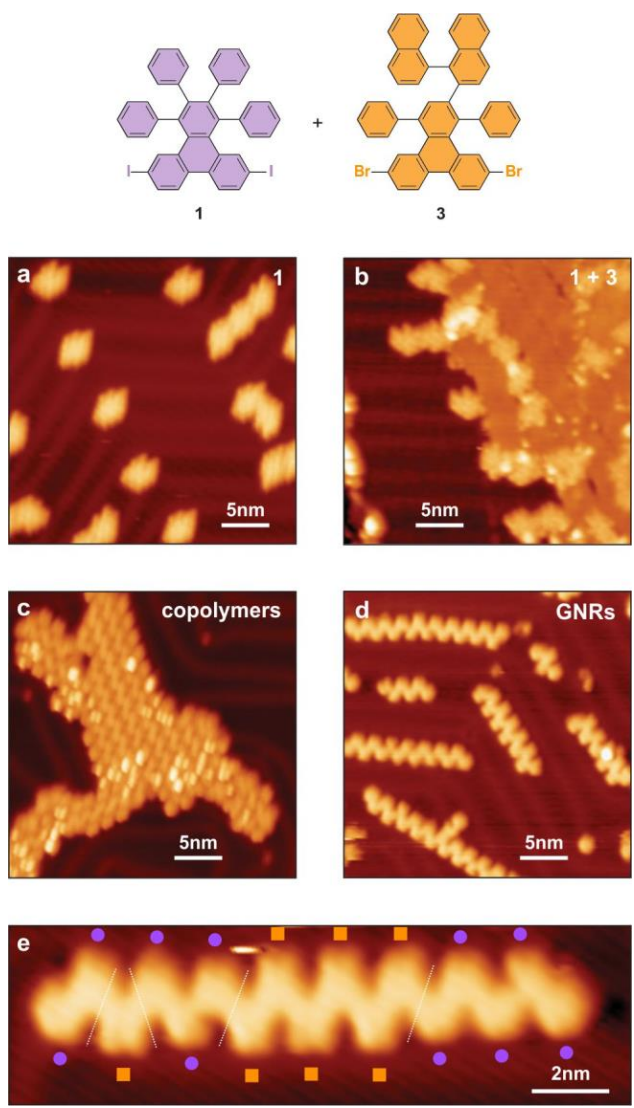

**Supplementary Figure S1 | Control experiment omitting the linker molecule. a**, STM image of **1** on Au(111) ($V =$ 1.0 V, $I =$ 20 pA). **b**, STM image of **1** and **3** co-adsorbed on Au(111) ($V =$ 2.0 V, $I =$ 20 pA). **c**, Copolymer island after annealing to 230 °C ($V =$ 1.0 V, $I =$ 20 pA). **d**, GNRs observed after annealing to 320 °C ($V =$ 0.3 V, $I =$ 40 pA). **e**, Magnified GNR exhibiting four heterojunctions ($V =$ 0.3 V, $I =$ 40 pA). Purple round markers indicate cGNR segments, orange square markers indicate binaph-cGNR segments.

As described in the main text, two control experiments were conducted to assess the impact of the hierarchical growth strategy on the control of the copolymer growth sequence. The first control experiment is shown in Supplementary Fig. S1. Monomers **1** and **3** were deposited onto a clean Au(111) surface (Supplementary Figs. S1a,b) and annealed at 230 °C resulting in islands of copolymers in which binaphthyl segments are evident as taller protrusions (Supplementary Fig. 1c). Further annealing at 320

°C induced cyclodehydrogenation (Supplementary Figs. 1d,e). Fig. 4c in the main article shows that compared to the full hierarchical protocol using precursors **1–3**, significantly fewer GNRs possess only a single heterojunction, indicating that the linker molecule **2** (omitted in this first control experiment) increases the degree of control over the growth sequence.

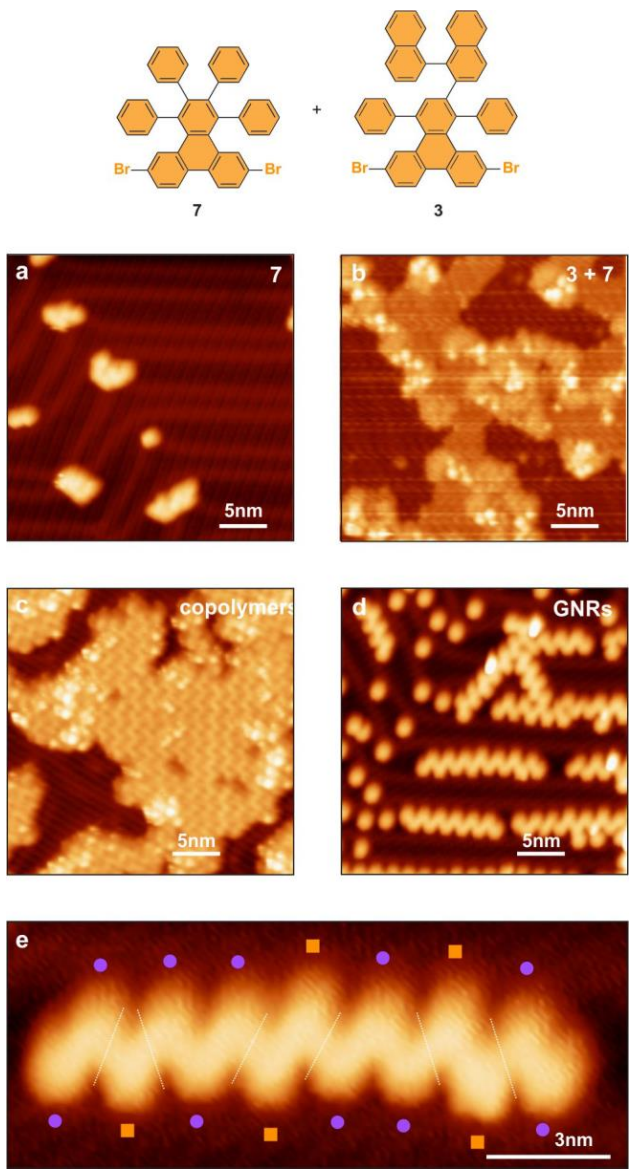

**Supplementary Figure S2 | Control experiment using only brominated precursors. a**, STM image of **7** deposited on Au(111) ($V$ = 1.0 V, $I$ = 30 pA). **b**, STM image of **7** and **3** co-deposited on Au(111) ($V$ = 1.0 V, $I$ = 20 pA). **c**, Copolymer island after annealing to 190 °C (V = 1.0 V, $I$ = 20 pA). **d**, STM image of GNRs obtained after annealing to 330 °C ($V$ = 0.3 V, $I$ = 40 pA). **e**, Magnified image of GNR with six heterojunctions ($V$ = 0.3 V, $I$ = 40 pA). Purple round markers indicate cGNR segments, orange square markers indicate binaph-cGNR segments.

The second control experiment shown in Supplementary Fig. S2 uses only brominated precursors, namely **7** and **3**. After sequential deposition (Supplementary Figs. S2a,b) of the monomers on Au(111), they were annealed at 190 °C resulting in islands of copolymers (Supplementary Fig. S2c). Further annealing at 330 °C leads to cyclodehydrogenation and yields GNRs (Supplementary Figs. S2d,e). Fig. 4c in the main article shows that using precursors **7** and **3**, there is a higher relative occurrence of GNRs with more than four heterojunctions and a dramatically reduced occurrence of single-junction GNRs compared to both the preparation using **1** and **3** (partial hierarchical protocol) and the full hierarchical protocol using **1**–**3**.

The comparison between the two control experiments demonstrates that substituting one of the monomers with iodine significantly increases the control over the growth sequence as indicated by the number of heterojunctions in a given GNR. The two control experiments combined demonstrate that the hierarchical growth strategy as implemented in the full hierarchical protocol using **1**–**3** results in an increase of control in GNR heterojunction synthesis.

## 2. Length Distribution of GNRs

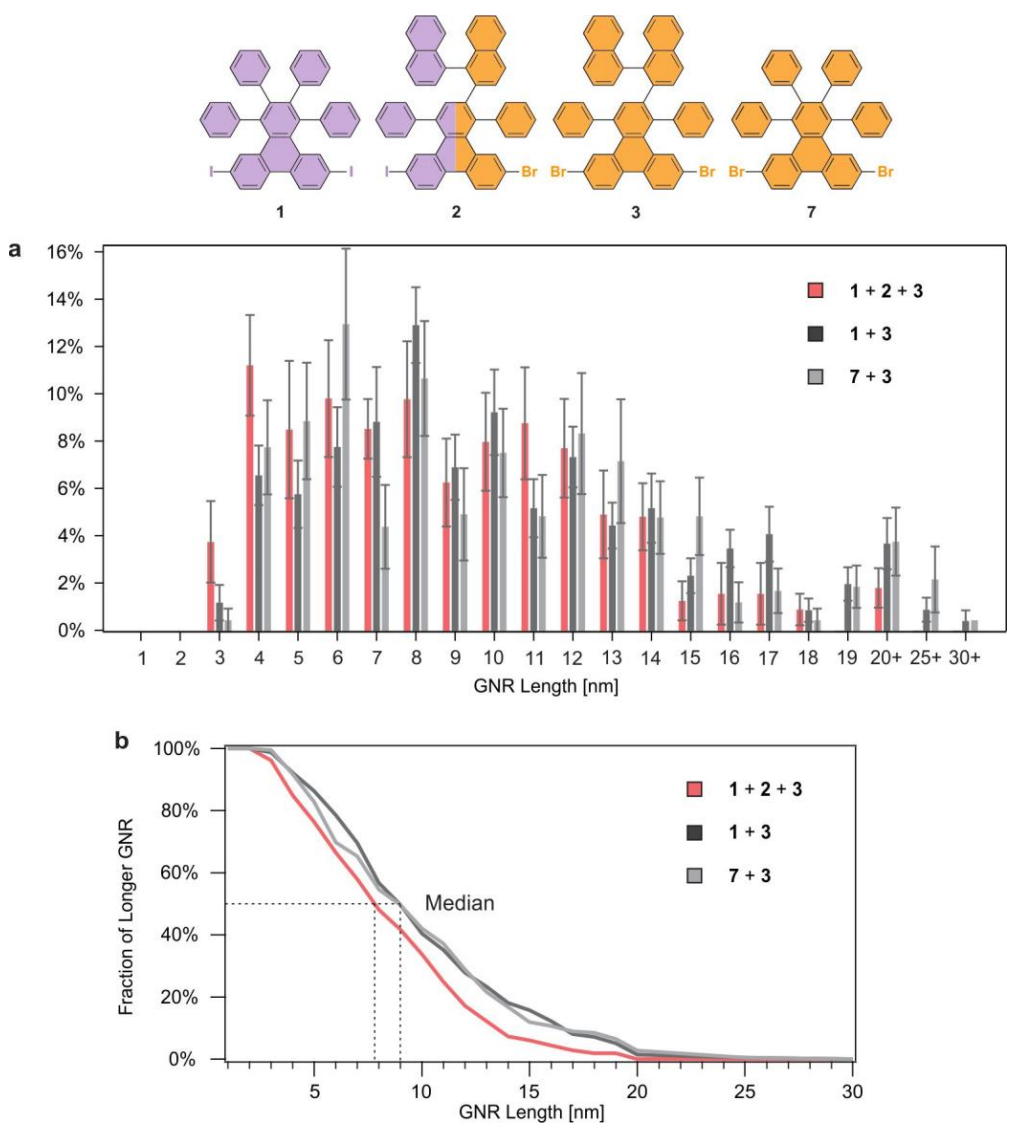

**Supplementary Figure S3 | Length distribution of GNRs. a**, Length histograms for the three different heterojunction preparations on Au(111) using the three different growth protocols. **b**, Integrated histogram for the three different growth protocols showing the fraction of GNRs longer than a given length. The median GNR length can be read as indicated.

Supplementary Fig. S3a shows the length distribution of the GNR heterostructures that were obtained in the three different growth protocols using different combinations of precursors, namely using precursors **1–3** (full hierarchical protocol), **1** and **3** (partial hierarchical protocol) and **7** and **3** (random protocol). No significant differences in the length distribution can be observed for the three cases. The majority of GNRs possesses a length between 4 and 20 nm. In Supplementary Fig. S3b, the histograms were integrated to show the fraction of GNRs which exceed a given length. From this diagram, the median length can be directly obtained. In all three cases the median length is 8 or 9 nm.

### 3. Original STM image of GNR Heterostructures

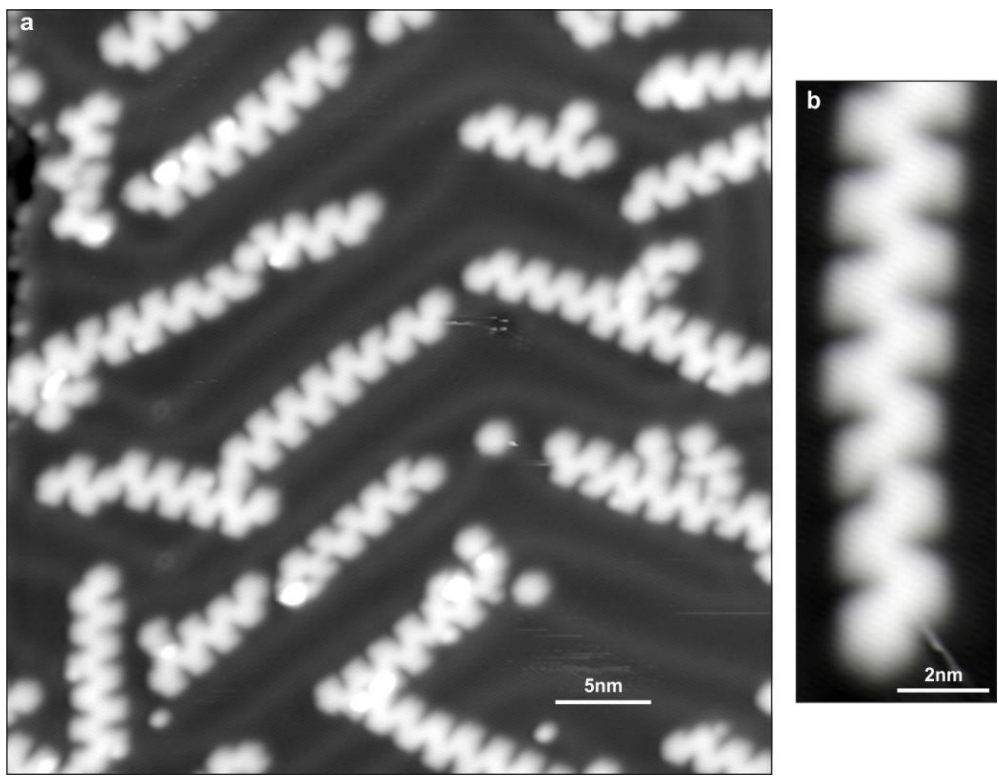

**Supplementary Figure S4 | Original STM images of GNR heterostructures. a**, STM image of GNR heterostructures shown in Fig. 4a without overlaid tinting ($V$ = 0.3 V, $I$ = 20 pA). **b**, STM image of the GNR heterostructure shown in Fig. 4b without overlaid tinting ($V$ = 0.3 V, $I$ = 20 pA).

The STM images of GNR heterostructures in the main article were tinted to help identify cGNR and binaph-cGNR segments. Supplementary Fig. S4 shows the original, unprocessed images that were used in Figs. 4a,b.

## 4. Calculated Projected Density of States (PDOS) of GNR Heterojunction

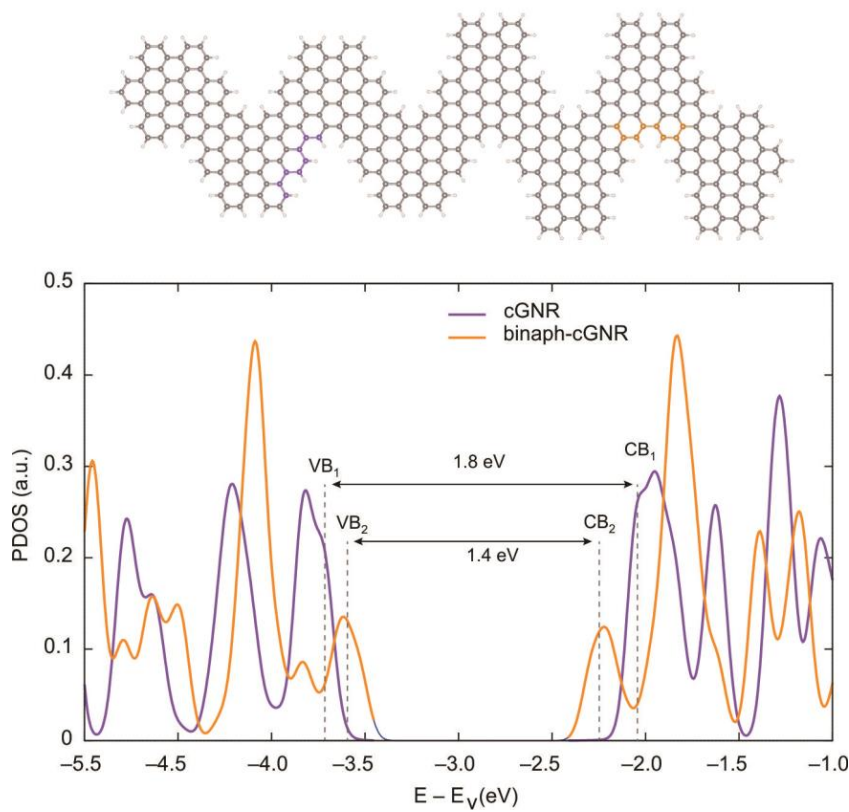

**Supplementary Figure S5 | Electronic structure of cGNR and binaph-cGNR segments from DFT calculations.** PDOS averaged in the cGNR (purple) and binaph-cGNR (orange) segment of a single molecular GNR heterojunction. The horizontal axis denotes the energy relative to the vacuum level. The band gap energies of cGNR and binaph-cGNR are given by 1.8 eV and 1.4 eV, respectively.

We have calculated the projected density of states (PDOS) using the Kohn-Sham states within LDA-DFT for a cGNR/binaph-cGNR heterojunction (the unit cell is shown in Supplementary Fig. S5) in order to compare with the differential conductance spectra in Fig. 5 of the main text. To mimic the broadening of the STM tip, the PDOS is obtained over an average of eight carbon atoms in each area (as highlighted in purple and orange for cGNR and binaph-cGNR, respectively (Supplementary Fig. S5, top)). The simulated PDOS spectral density (Supplementary Fig. S5) is qualitatively similar to the experimental STS results (including a particularly large PDOS intensity for the CB+1 band in the binaphthyl segment). In the LDA-DFT calculations the gap energies of the cGNR and binaph-cGNR segments in the heterojunction are given by 1.8 eV and 1.4 eV. If we include electron correlation effects to the self energy of the electron states within the GW approximation, the band gap increases to the large values (3.63 eV for a cGNR). Furthermore, if we included the screening from the gold substrate, the band gap energy would be reduced again similar to the former study on 7-13 AGNR heterojunctions.[1] These two effects account for the difference of the band gap magnitude between the LDA calculations and the STS experiments.

## 5. Synthesis of Molecular Precursors

**Materials and General Methods.** Unless otherwise stated, all manipulations of air and/or moisture sensitive compounds were carried out in oven-dried glassware, under an atmosphere of $N_2$. All solvents and reagents were purchased from Alfa Aesar, Spectrum Chemicals, Acros Organics, TCI America, and Sigma-Aldrich and were used as received unless otherwise noted. Organic solvents were dried by passing through a column of alumina and were degassed by vigorous bubbling of $N_2$ through the solvent for 20 min. Flash column chromatography was performed on SiliCycle silica gel (particle size 40–63 μm). Thin layer chromatography was carried out using SiliCycle silica gel 60 Å F-254 precoated plates (0.25 mm thick) and visualized by UV absorption. All $^1H$ and $\{^1H\}^{13}C$ NMR spectra were recorded on Bruker AV-600, AV-500, and AVQ-400 spectrometers, and are referenced to residual solvent peaks ($C_2D_2Cl_4$ $^1H$ NMR $\delta$ = 6.0 ppm, $^{13}C$ NMR $\delta$ = 73.8 ppm; $CDCl_3$ $^1H$ NMR $\delta$ = 7.26 ppm, $^{13}C$ NMR $\delta$ = 77.2 ppm; $(CD_3)_2SO$ $^1H$ NMR $\delta$ = 2.50, $^{13}C$ NMR $\delta$ = 39.5); ESI-HRMS mass spectrometry was performed on a Finnigan LTQ FT (Thermo) via direct injection using a flow rate of 5.0 μL min$^{–1}$. X-ray quality single crystals of **2** were obtained by recrystallization from a MeOH/CHCl$_3$ solution. X-ray crystallography of **2** was performed on an APEX II QUAZAR, using a Microfocus Sealed Source (Incoatec; Cu-K$_\alpha$ radiation), Kappa Geometry with DX (Bruker-AXS build) goniostat, a Bruker APEX II detector, QUAZAR multilayer mirrors as the radiation monochromator, and Oxford Cryostream 700 held at 100 K. Crystallographic data were resolved with SHELXT, refined with SHELXL-2014, and visualized with ORTEP-32. Compounds 6,11-diiodo-1,2,3,4-tetraphenyltriphenylene (**1**), 6,11-dibromo-1,2,3,4-tetraphenyltriphenylene (**7**), and 2,7-dibromophenanthrene-9,10-dione were synthesized following previously reported literature procedures.[2,3]

**2-Bromo-1,1'-binaphthyl (8):** A 50 mL 2-neck round bottom flask was charged with 2,2'-dibromo-1,1'-dinaphthyl (0.98 g, 2.4 mmol) and dry THF (15 mL). The solution was cooled to −40 °C, $^n$BuLi (2.5 M in hexanes, 2.4 mmol) was added dropwise and the reaction mixture was stirred for 1 h at −40 °C. The reaction mixture was cooled to −78 °C, HCl (1 N, 11.8 mmol) added dropwise, and the reaction mixture was warmed to 25 °C. The solution was concentrated on a rotary evaporator. The residue was dissolved in Et$_2$O (50 mL) and hydrolyzed with 1 N HCl (50 mL). The reaction mixture was extracted with Et$_2$O, the combined organic layers were washed with saturated aqueous NaHCO$_3$ solution, water, and saturated aqueous NaCl solution, and dried over MgSO$_4$. The reaction mixture was concentrated on a rotary evaporator to give **8** (0.72 g, 92%) as a colorless solid. Spectroscopic data is consistent with literature reports.[4] $^1H$ NMR (400 MHz, 25 °C, CDCl$_3$) $\delta$ = 8.09 (dd, *J* = 13.4, 8.3 Hz, 2 H), 7.99 (d, *J* = 8.2 Hz, 1H), 7.96–7.87 (m, 2H), 7.74 (t, *J* = 8.1 Hz, 1H), 7.65–7.48 (m, 3H), 7.46–7.38 (m, 2H), 7.38–7.28 (m, 2H) ppm.

**2-ethynyl-1,1'-binaphthalene (9):** A 25 mL sealable Schlenk flask was charged with **8** (305 mg, 0.75 mmol) in diisopropyl amine (11 mL) and THF (3 mL). Pd(PPh$_3$)$_4$ (106 mg, 0.08 mmol) and CuI (10 mg, 0.04 mmol) were added, and the reaction mixture was degassed. TMSA (2 mL) was added, and the flask was sealed and stirred at 55 °C for 20 h. The reaction mixture was cooled to 25 °C, and extracted with Et$_2$O. The combined organic layers were dried over MgSO$_4$, and concentrated on a rotary evaporator. Column chromatography (hexanes) yielded an inseparable mixture of partially deprotected **9**. The intermediate was redissolved in THF (8 mL) and MeOH (8 mL). K$_2$CO$_3$ (1 g) was added, and the reaction mixture was stirred at 25 °C for 2 h. The reaction mixture was extracted with CH$_2$Cl$_2$, and the combined organic layers were washed with H$_2$O, dried over MgSO$_4$, and concentrated on a rotatory evaporator. Column chromatography (hexane/CH$_2$Cl$_2$ 1:0–10:1) yielded **9** (170 mg, 81%) as a colorless oil. Spectroscopic data is consistent with literature reports.[5] $^1$H NMR (400 MHz, 25 °C, CDCl$_3$) $\delta$ = 8.03–7.86 (m, 4H), 7.71 (d, $J$ = 8.6 Hz, 1H), 7.65–7.58 (m, 1H), 7.53–7.43 (m, 3H), 7.33–7.17 (m, 4H), 2.8 (s, 1H) ppm.

**2-([1,1'-binaphthalen]-2-yl)-6,11-dibromo-1,4-diphenyltriphenylene (3):** A 2-neck 50 mL round bottom flask was charged with 2,7-dibromophenanthrene-9,10-dione (1.0 g, 2.7 mmol) in MeOH (5.7 mL). Diphenyl acetone (5.1 g, 3.6 mmol) was added and the reaction mixture was heated to 75 °C. KOH (0.16 g, 2.9 mmol) in MeOH (9.8 mL) was added drop-wise, and the reaction mixture was heated at 75 °C for 2 h. The reaction was cooled to 25 °C, and filtered. The precipitate was washed with EtOH to yield the intermediate cyclopentadienone as a green solid, which was used without further purification. A 10 mL Schlenk flask was charged with the crude cyclopentadienone (19 mg, 0.036 mmol) and **9** (11 mg, 0.039 mmol) in Ph$_2$O (0.7 mL), and the reaction mixture was heated to 145 °C for 16 h. The reaction mixture was cooled to 25 °C, and the solvent was evaporated to yield a crude residue. Column chromatography (CH$_2$Cl$_2$/hexane 1:10) yielded **3** (21 mg, 76%) as a colorless solid. Variable temperature NMR in 1,1,2,2-tetrachloroethane-$d_2$ at 110 °C was performed in order to resolve NMR spectroscopic signals (Supplementary Fig. S6). Even at high temperature, a fully resolved NMR was not obtained due to the high barrier to rotation within the molecule. Major NMR shifts at 22 °C reported as follows (Supplementary Figs. S8, S9). $^1$H NMR (600 MHz, CDCl$_3$) $\delta$ = 8.13 (d, $J$ = 8.7 Hz, 1H), 7.93 (m, 1H), 7.91 (m, 1H), 7.70–7.64 (m, 2H), 7.47–7.38 (m, 4H), 7.26–7.09 (m, 3H), 6.73 (s, 1H) ppm; $^{13}$C NMR (151 MHz, CDCl$_3$) $\delta$ = 143.5, 141.6, 138.6, 137.7, 137.2, 133.7, 133.5, 132.8, 131.7, 130.2, 129.9, 129.7, 129.6, 129.2, 128.0, 127.6, 127.6, 127.2, 127.0 (2C), 126.6, 126.2 (2C), 126.1, 125.7, 125.4, 124.7, 124.6 120.1 ppm. HRMS (EI) $m/z$: [C$_{50}$H$_{30}$Br$_2$]$^+$, calcd. for [C$_{50}$H$_{30}$Br$_2$] 790.0681; found 790.0694.

**2-iodophenanthrene-9,10-dione (10):** A 5 mL sealable flask was charged with 9,10-phenanthrene quinone (0.25 g, 1.2 mmol) and cooled to 0 °C. Trifluoromethanesulfonic acid (1.0 mL) was added, and the reaction mixtures was stirred under $N_2$ for 10 min. *N*-Iodosuccinimide (0.54 g, 2.4 mmol) was added slowly to the suspension. The reaction was poured onto ice (100 mL) and the precipitate was filtered to yield **10** (0.07 g, 18%) as an orange solid. Spectroscopic data is consistent with literature reports.[6] $^1$H NMR (400 MHz, 25 °C, DMSO-$d_6$) $\delta$ = 8.36–8.20 (m, 2H), 8.10–8.03 (m, 3H), 7.78 (t, *J* = 7.7 Hz, 1H), 7.56 (t, 7.5 Hz, 1H) ppm.

**2-bromo-7-iodophenanthrene-9,10-dione (11):** A 25 mL 3-neck round bottom flask was charged with **10** (50 mg, 0.15 mmol) in $H_2SO_4$ (1.5 mL). *N*-Bromosuccinimide (29 mg, 0.16 mmol) was added, and the reaction was stirred at 25 °C for 16 h. The reaction was poured over ice (100 mL), and the precipitate was filtered to yield **11** (43 mg, 70%) as an orange solid. $^1$H NMR (600 MHz, 25 °C, DMSO-$d_6$) $\delta$ = 8.31–8.21 (m, 2H), 8.18–8.03 (m, 3H), 8.00–7.89 (m, 1H) ppm. HRMS (EI) *m/z*: $[C_{14}H_6O_2BrI]^+$, calcd. for $[C_{14}H_6O_2BrI]$ 411.8601; found 411.8596.

**2-([1,1'-binaphthalen]-2-yl)-6-bromo-11-iodo-1,4-diphenyltriphenylene (2):** A 2-neck 25 mL round bottom flask was charged with **11** (83 mg, 0.20 mmol) in MeOH (2 mL). 1,3-diphenyl acetone (55 mg, 0.26 mmol) was added and the reaction mixture was heated to 75 °C. KOH (12 mg, 0.21 mmol) in MeOH (4 mL) was added drop-wise, and the reaction mixture was heated at 75 °C for 2 h. The reaction was cooled to 25 °C, and filtered. The precipitate was washed with EtOH to yield the intermediate cyclopentadienone as a green solid, which was used without further purification. A 10 mL Schlenk flask was charged with the crude cyclopentadienone (30 mg, 0.051 mmol) and **9** (16 mg, 0.056 mmol) in *o*-xylene (1 mL), and the reaction mixture was heated to 145 °C for 16 h. The reaction mixture was cooled to 25 °C and the solvent was evaporated to yield a crude residue. Column chromatography ($CH_2Cl_2$/hexane 1:10) to yield **2** (20 mg, 47%) as a colorless solid. Variable temperature NMR in 1,1,2,2-tetrachloroethane-$d_2$ at 110 °C was performed in order to resolve NMR spectroscopic signals (Supplementary Fig. S7). Even at high temperature, a fully resolved NMR was not obtained due to the high degree of rotational restriction within the molecule. Major NMR shifts at 20 °C reported as follows (Supplementary Figs. S10, S11). $^1$H NMR (600 MHz, CDCl$_3$) $\delta$ = 8.13 (d, *J* = 8.7 Hz, 1H), 7.98 (d, *J* = 8.5 Hz, 1H), 7.91 (m, 2H), 7.85 (s, 1H), 7.80 (d, *J* = 8.4 Hz, 1H), 7.71–7.54 (m, 4H), 7.54–7.34 (m, 6H), 7.24–7.18 (m, 3H), 7.18–6.89 (m, 3H), 6.55–6.39 (m, 2H) ppm; $^{13}$C NMR (151 MHz, 1,1,2,2-tetrachloroethane-$d_2$) $\delta$ = 142.9, 141.0, 139.4, 138.8, 138.2, 137.1, 136.7, 136.0, 135.7, 134.9, 133.2, 132.9, 132.7, 132.5, 132.3, 132.3, 129.3, 129.2, 128.8, 128.2, 128.0, 127.8, 127.6, 127.2, 127.2, 126.7, 126.6, 126.4, 126.0, 125.7,

125.7, 125.3, 124.9, 124.2, 124.1, 124.0 ppm. HRMS (EI) *m/z*: [C$_{50}$H$_{30}$BrI]$^+$, calcd. for [C$_{50}$H$_{30}$BrI] 838.0563; found 838.0555.

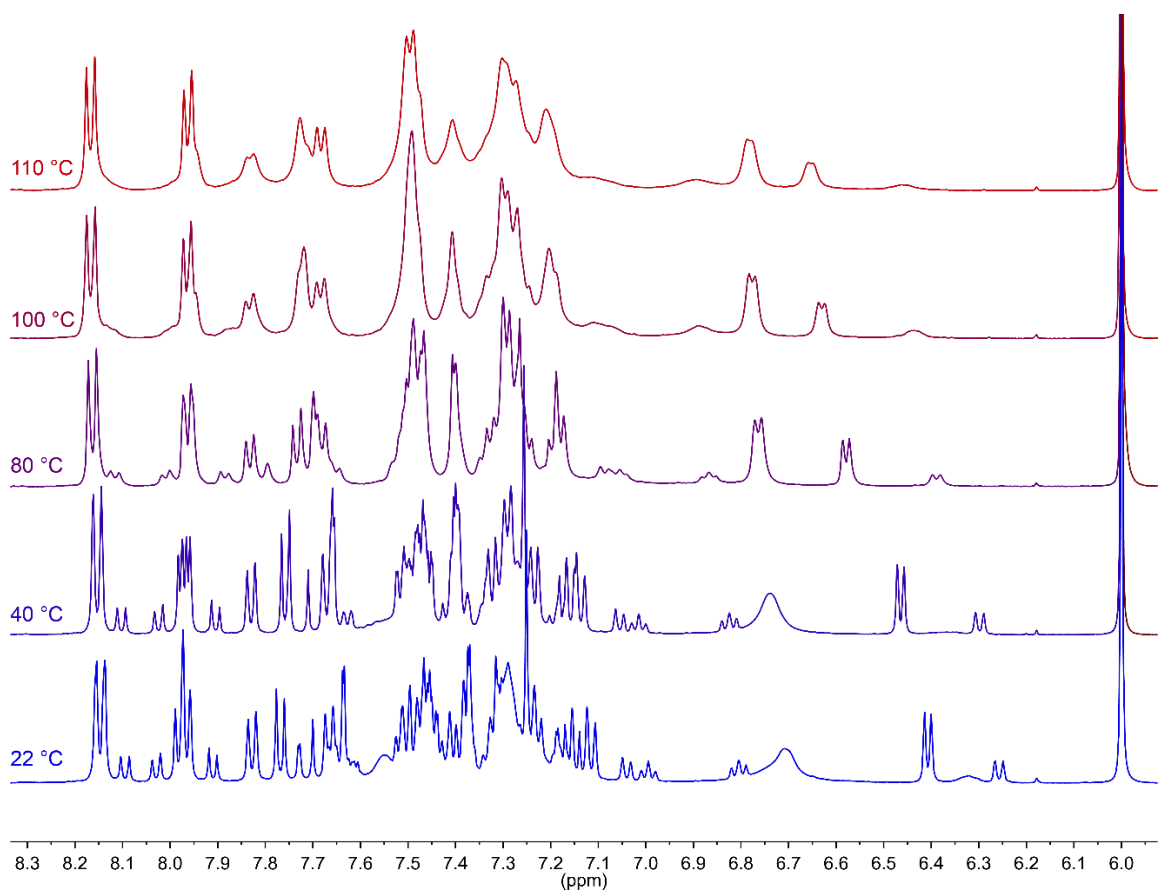

**Supplementary Figure S6 |** Variable Temperature (22–110 °C) NMR of **3**, $^1$H NMR (500 MHz, 1,1,2,2-tetrachloroethane-$d_2$).

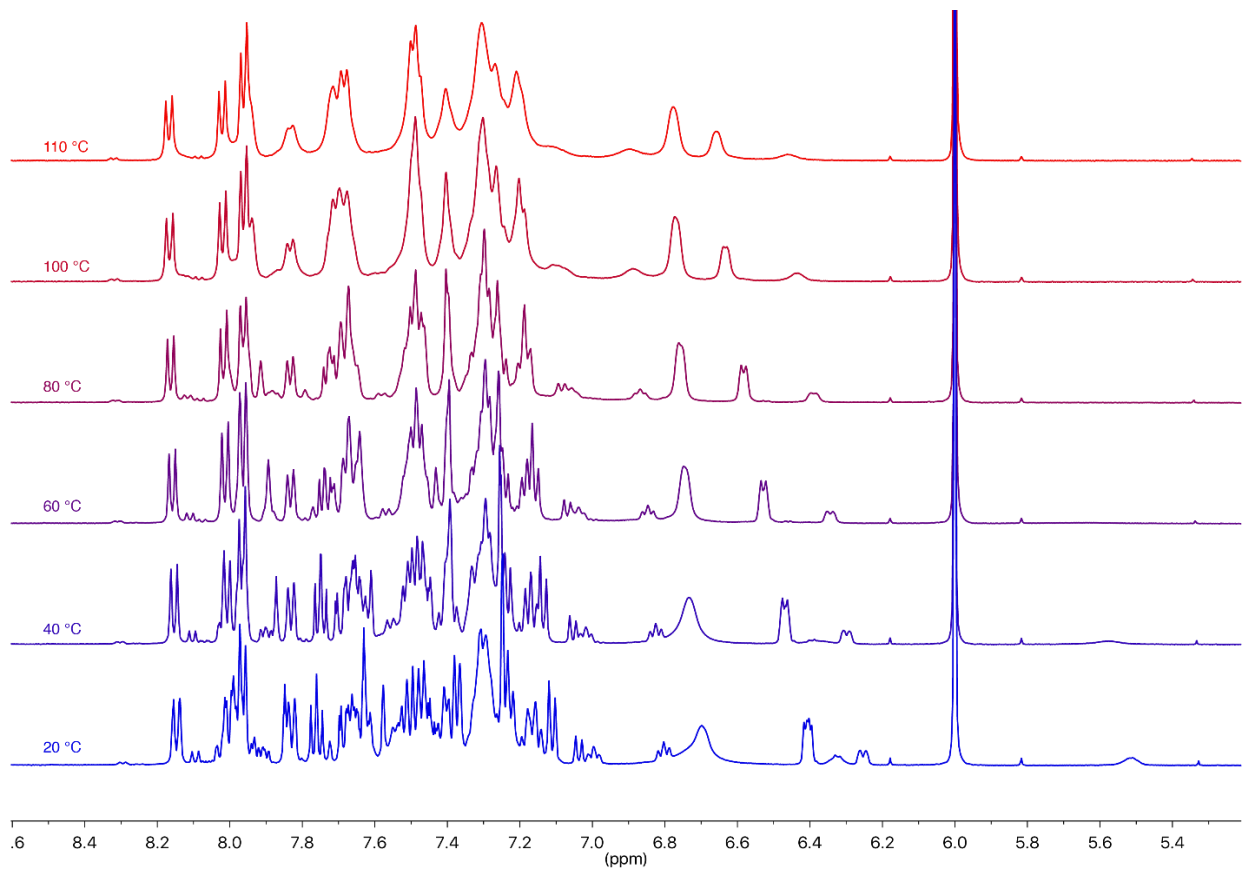

**Supplementary Figure S7 |** Variable Temperature (20–110 °C) NMR of **2**, [1]H NMR (500 MHz, 1,1,2,2-tetrachloroethane-$d_2$).

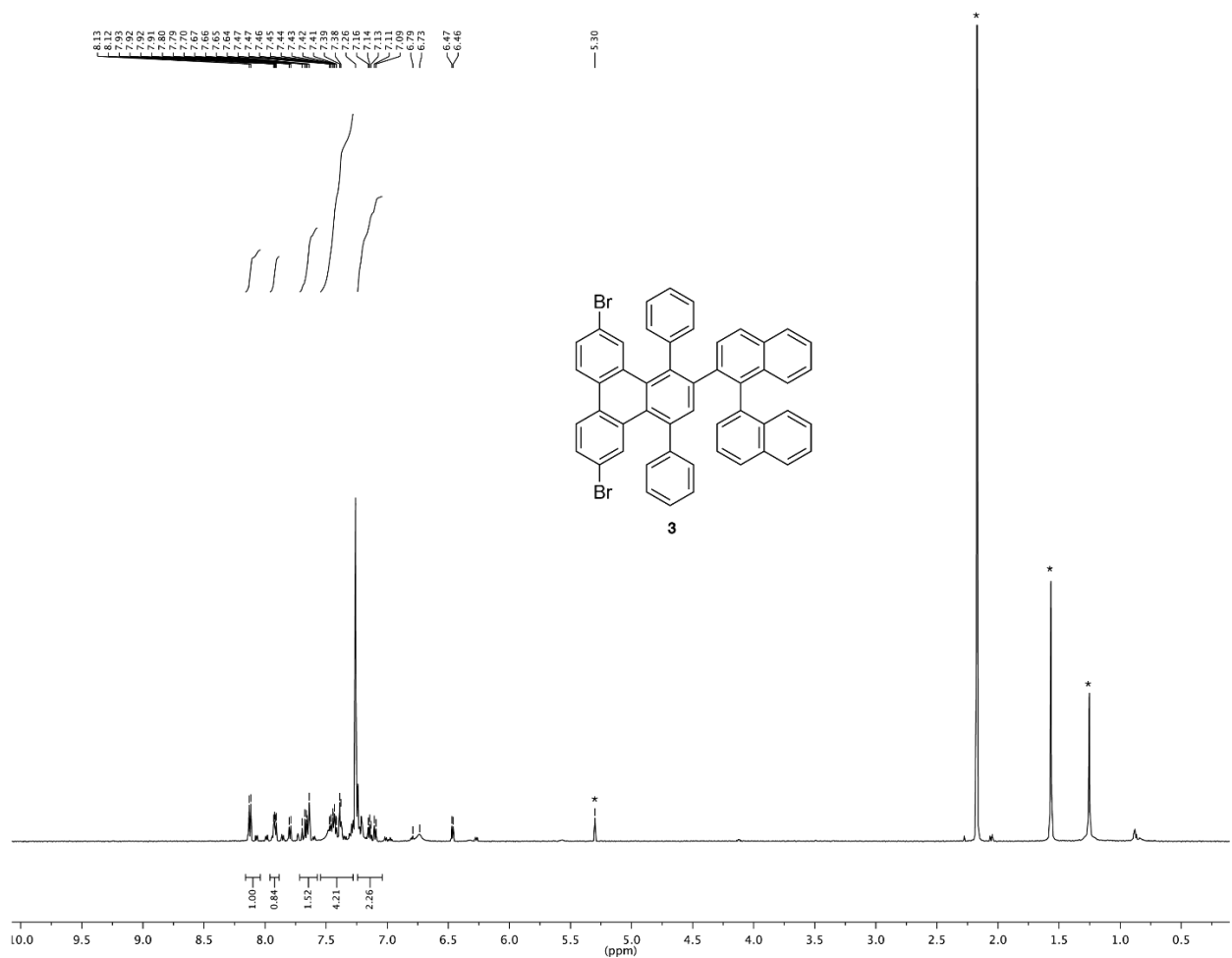

**Supplementary Figure S8 |** $^1$H NMR (600 MHz, 20 °C, CDCl$_3$) of **3**; [ * ] indicate residual solvent signals.

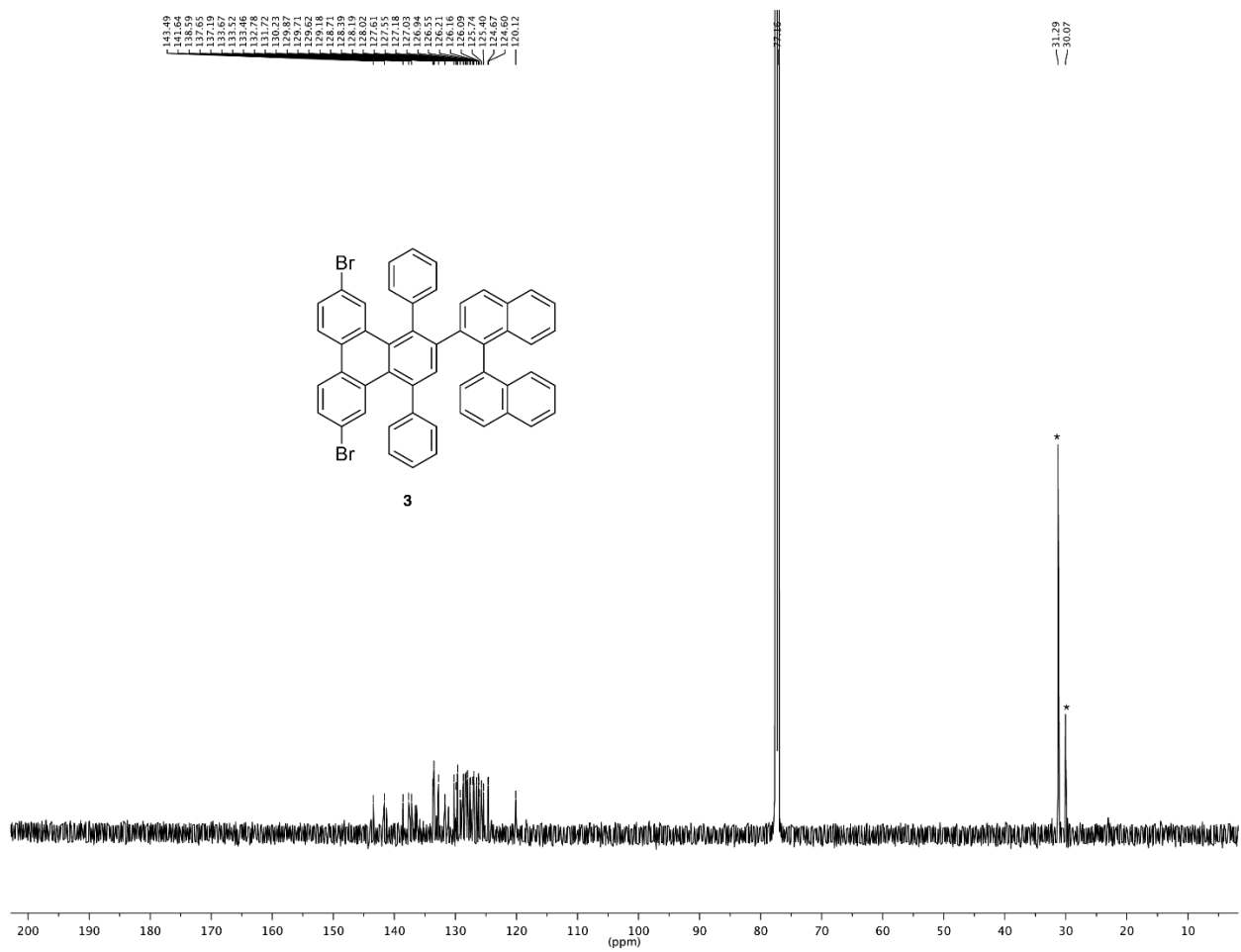

**Supplementary Figure S9 |** $^{13}$C NMR (151 MHz, 20 °C, CDCl$_3$) of **3**; [ * ] indicate residual solvent signals.

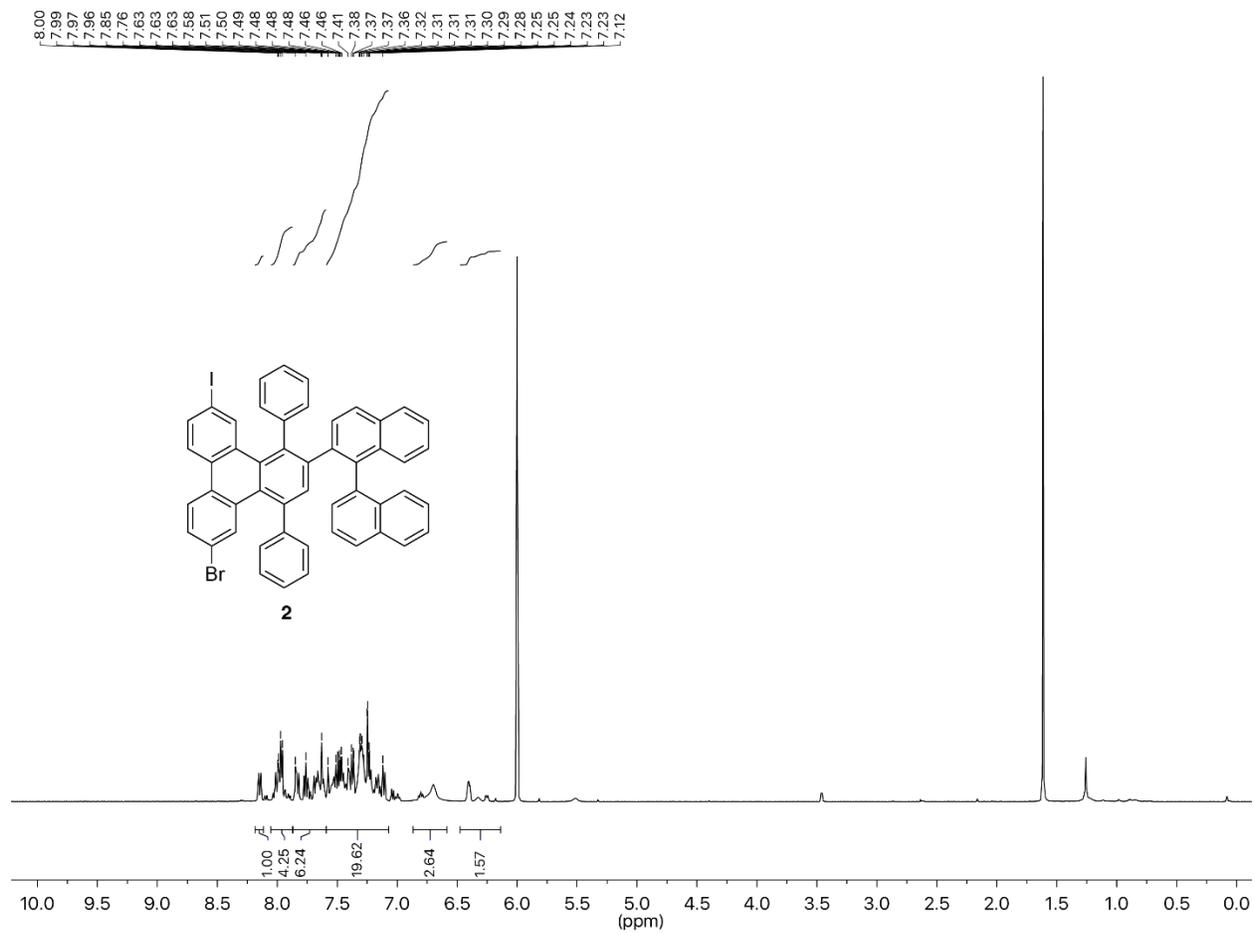

**Supplementary Figure S10 |** $^1$H NMR (500 MHz, 20 °C, 1,1,2,2-tetrachloroethane-$d_2$) of **2**; [ * ] indicate residual solvent signals.

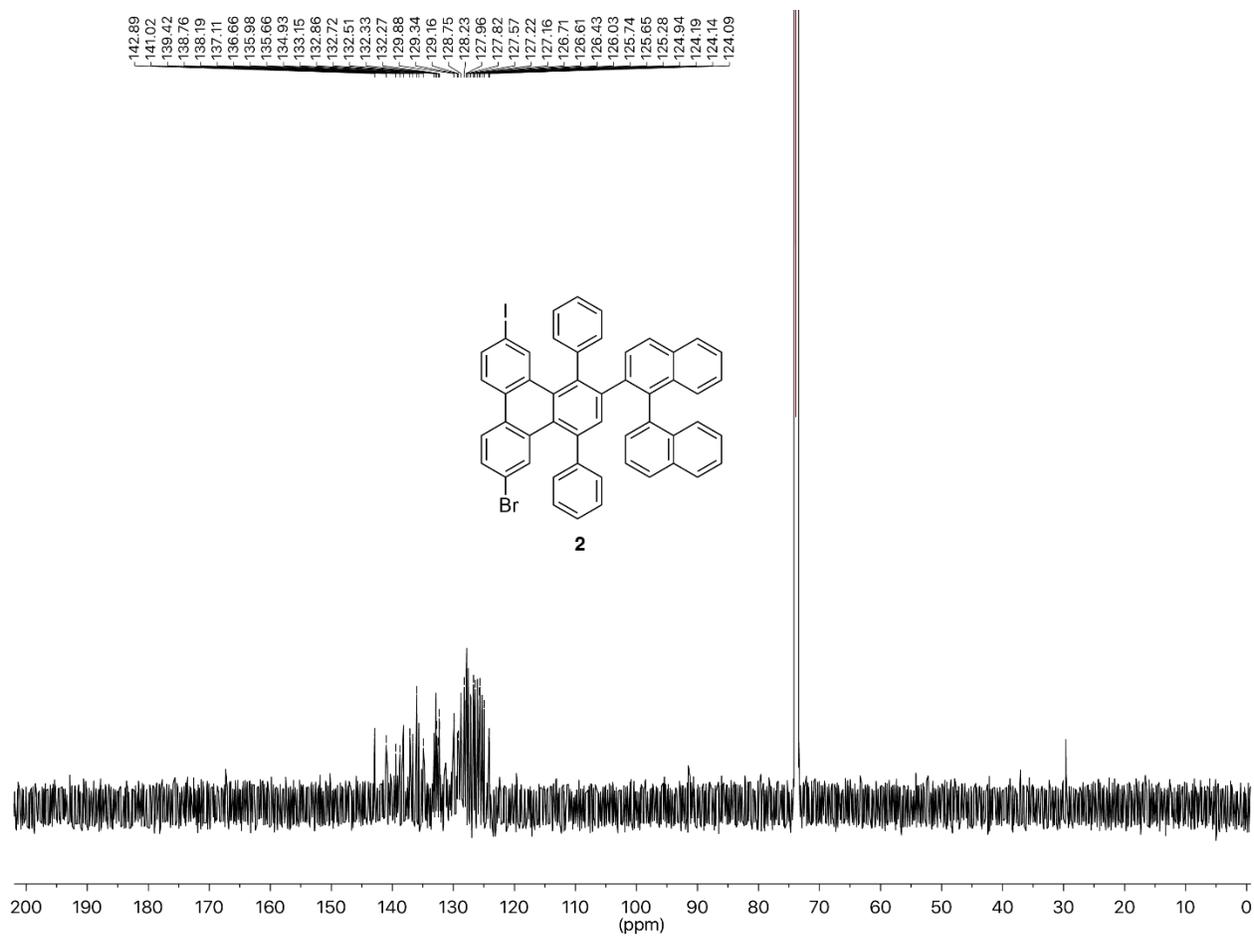

**Supplementary Figure S11 |** $^{13}$C NMR (151 MHz, 20 °C, 1,1,2,2-tetrachloroethane-$d_2$) of **2**; [ * ] indicate residual solvent signals.

**Supplementary Table S1 |** Crystal data and structure refinement for **2**

| | |
|---|---|
| CDCC no. | 1568277 |
| Empirical formula | $C_{50}H_{30}BrI$ |
| Formula weight | 837.55 |
| Temperature | 100(2) K |
| Wavelength | 0.71073 Å |
| Crystal system | Triclinic |
| Space group | P–1 |
| Unit cell dimensions | $a$ = 11.4073(4) Å     $\alpha$ = 113.8930(10)° |
| | $b$ = 12.4337(5) Å     $\beta$ = 91.728(2)° |
| | $c$ = 13.8759(5) Å     $\gamma$ = 93.409(2)° |
| Volume | 1793.07(12) Å$^3$ |
| Z | 2 |
| Density (calculated) | 1.551 Mg/m$^3$ |
| Absorption coefficient | 2.044 mm$^{-1}$ |
| F(000) | 836 |
| Crystal size | 0.070 x 0.050 x 0.050 mm$^3$ |
| Theta range for data collection | 1.608 to 25.457°. |
| Index ranges | $-13 \leq h \leq 13, -14 \leq k \leq 15, -16 \leq l \leq 16$ |
| Reflections collected | 97222 |
| Independent reflections | 6581 [R(int) = 0.0414] |
| Completeness to theta = 25.000° | 99.9 % |
| Absorption correction | Semi-empirical from equivalents |
| Max. and min. transmission | 0.745 and 0.643 |
| Refinement method | Full-matrix least-squares on F$^2$ |
| Data / restraints / parameters | 6581 / 0 / 476 |
| Goodness-of-fit on F$^2$ | 1.086 |
| Final R indices [I>2sigma(I)] | R1 = 0.0254, wR2 = 0.0579 |
| R indices (all data) | R1 = 0.0315, wR2 = 0.0622 |
| Extinction coefficient | n/a |
| Largest diff. peak and hole | 0.514 and –0.541 e Å$^{-3}$ |

**Supplementary Table S2 |** Atomic coordinates (x 10⁴) and equivalent isotropic displacement parameters (Å² x 10³) for **3**. U(eq) is defined as one third of the trace of the orthogonalized $U^{ij}$ tensor.

_________________________________________________________________________________

|       | x       | y       | z       | U(eq) |
|-------|---------|---------|---------|-------|
| C(1)  | 4247(2) | 5366(2) | 6741(2) | 15(1) |
| C(2)  | 5406(2) | 5749(2) | 7140(2) | 16(1) |
| C(3)  | 6269(2) | 4980(2) | 7050(2) | 15(1) |
| C(4)  | 5979(2) | 3751(2) | 6464(2) | 15(1) |
| C(5)  | 6731(2) | 2827(2) | 6460(2) | 16(1) |
| C(6)  | 7571(2) | 3042(2) | 7294(2) | 19(1) |
| C(7)  | 8246(2) | 2166(2) | 7301(2) | 23(1) |
| C(8)  | 8121(2) | 1037(2) | 6491(2) | 26(1) |
| C(9)  | 7264(2) | 798(2)  | 5694(2) | 25(1) |
| C(10) | 6544(2) | 1668(2) | 5668(2) | 20(1) |
| C(11) | 5607(2) | 1408(2) | 4848(2) | 19(1) |
| C(12) | 5541(2) | 374(2)  | 3907(2) | 25(1) |
| C(13) | 4705(2) | 169(2)  | 3104(2) | 27(1) |
| C(14) | 3930(2) | 1027(2) | 3220(2) | 22(1) |
| C(15) | 3949(2) | 2039(2) | 4130(2) | 19(1) |
| C(16) | 4767(2) | 2237(2) | 4977(2) | 17(1) |
| C(17) | 4876(2) | 3385(2) | 5888(2) | 16(1) |
| C(18) | 3958(2) | 4171(2) | 6133(2) | 15(1) |
| C(19) | 3399(2) | 6306(2) | 6976(2) | 15(1) |
| C(20) | 3041(2) | 6635(2) | 6151(2) | 19(1) |
| C(21) | 2343(2) | 7537(2) | 6331(2) | 22(1) |
| C(22) | 1980(2) | 8202(2) | 7348(2) | 20(1) |
| C(23) | 1253(2) | 9150(2) | 7558(2) | 29(1) |
| C(24) | 934(2)  | 9790(2) | 8558(2) | 32(1) |
| C(25) | 1330(2) | 9527(2) | 9394(2) | 29(1) |
| C(26) | 2022(2) | 8612(2) | 9227(2) | 22(1) |
| C(27) | 2360(2) | 7916(2) | 8196(2) | 17(1) |
| C(28) | 3061(2) | 6934(2) | 7990(2) | 16(1) |
| C(29) | 3363(2) | 6554(2) | 8858(2) | 16(1) |
| C(30) | 4289(2) | 7162(2) | 9636(2) | 17(1) |
| C(31) | 4975(2) | 8160(2) | 9649(2) | 21(1) |
| C(32) | 5896(2) | 8681(2) | 10381(2) | 26(1) |
| C(33) | 6191(2) | 8232(2) | 11133(2) | 29(1) |
| C(34) | 5532(2) | 7295(2) | 11159(2) | 27(1) |
| C(35) | 4562(2) | 6738(2) | 10423(2) | 21(1) |
| C(36) | 3877(2) | 5756(2) | 10438(2) | 23(1) |
| C(37) | 2975(2) | 5208(2) | 9701(2) | 22(1) |
| C(38) | 2725(2) | 5604(2) | 8898(2) | 18(1) |
| C(39) | 7458(2) | 5517(2) | 7545(2) | 17(1) |
| C(40) | 8478(2) | 5220(2) | 6992(2) | 20(1) |
| C(41) | 9569(2) | 5731(2) | 7471(2) | 26(1) |
| C(42) | 9654(2) | 6551(2) | 8516(2) | 31(1) |
| C(43) | 8653(2) | 6875(2) | 9069(2) | 26(1) |
| C(44) | 7558(2) | 6356(2) | 8585(2) | 20(1) |
| C(45) | 2702(2) | 3669(2) | 5885(2) | 17(1) |
| C(46) | 1887(2) | 3934(2) | 5257(2) | 24(1) |
| C(47) | 744(2) | 3408(2) | 5052(2) | 28(1) |
| C(48) | 395(2) | 2619(2) | 5481(2) | 31(1) |
| C(49) | 1187(2) | 2349(2) | 6108(2) | 29(1) |
| C(50) | 2329(2) | 2868(2) | 6307(2) | 22(1) |
| Br(1) | 9444(8) | 2520(7) | 8466(5) | 32(1) |
| I(1) | 2717(4) | 855(3) | 2022(3) | 28(1) |
| Br(1A) | 2834(6) | 720(5) | 2047(5) | 28(1) |
| I(1A) | 9394(5) | 2617(5) | 8601(3) | 32(1) |

________________________________________________________________________________

**Supplementary Table S3 |** Bond lengths [Å] and angles [°] for **2**.

_______________________________________________________________________

| | |
|---|---|
| C(1)-C(18) | 1.391(3) |
| C(1)-C(2) | 1.396(3) |
| C(1)-C(19) | 1.501(3) |
| C(2)-C(3) | 1.386(3) |
| C(2)-H(2) | 0.9500 |
| C(3)-C(4) | 1.421(3) |
| C(3)-C(39) | 1.492(3) |
| C(4)-C(17) | 1.416(3) |
| C(4)-C(5) | 1.473(3) |
| C(5)-C(6) | 1.407(3) |
| C(5)-C(10) | 1.413(3) |
| C(6)-C(7) | 1.374(3) |
| C(6)-H(6) | 0.9500 |
| C(7)-C(8) | 1.395(3) |
| C(7)-Br(1) | 1.974(9) |
| C(7)-I(1A) | 2.058(6) |
| C(8)-C(9) | 1.379(3) |
| C(8)-H(8) | 0.9500 |
| C(9)-C(10) | 1.407(3) |
| C(9)-H(9) | 0.9500 |
| C(10)-C(11) | 1.459(3) |
| C(11)-C(12) | 1.409(3) |
| C(11)-C(16) | 1.411(3) |
| C(12)-C(13) | 1.375(4) |
| C(12)-H(12) | 0.9500 |
| C(13)-C(14) | 1.388(3) |
| C(13)-H(13) | 0.9500 |
| C(14)-C(15) | 1.373(3) |
| C(14)-Br(1A) | 1.918(6) |
| C(14)-I(1) | 2.065(4) |
| C(15)-C(16) | 1.409(3) |
| C(15)-H(15) | 0.9500 |
| C(16)-C(17) | 1.470(3) |
| C(17)-C(18) | 1.429(3) |
| C(18)-C(45) | 1.500(3) |
| C(19)-C(28) | 1.385(3) |
| C(19)-C(20) | 1.419(3) |
| C(20)-C(21) | 1.358(3) |
| C(20)-H(20) | 0.9500 |
| C(21)-C(22) | 1.407(3) |
| C(21)-H(21) | 0.9500 |
| C(22)-C(23) | 1.419(3) |
| C(22)-C(27) | 1.423(3) |
| C(23)-C(24) | 1.365(4) |
| C(23)-H(23) | 0.9500 |
| C(24)-C(25) | 1.396(4) |
| C(24)-H(24) | 0.9500 |
| C(25)-C(26) | 1.370(3) |
| C(25)-H(25) | 0.9500 |
| C(26)-C(27) | 1.419(3) |
| C(26)-H(26) | 0.9500 |
| C(27)-C(28) | 1.434(3) |
| C(28)-C(29) | 1.500(3) |
| C(29)-C(38) | 1.372(3) |
| C(29)-C(30) | 1.424(3) |
| C(30)-C(31) | 1.420(3) |
| C(30)-C(35) | 1.425(3) |
| C(31)-C(32) | 1.370(3) |
| C(31)-H(31) | 0.9500 |
| C(32)-C(33) | 1.410(4) |
| C(32)-H(32) | 0.9500 |
| C(33)-C(34) | 1.363(4) |
| C(33)-H(33) | 0.9500 |
| C(34)-C(35) | 1.419(3) |
| C(34)-H(34) | 0.9500 |
| C(35)-C(36) | 1.417(3) |
| C(36)-C(37) | 1.362(3) |
| C(36)-H(36) | 0.9500 |
| C(37)-C(38) | 1.416(3) |
| C(37)-H(37) | 0.9500 |
| C(38)-H(38) | 0.9500 |
| C(39)-C(44) | 1.392(3) |
| C(39)-C(40) | 1.397(3) |
| C(40)-C(41) | 1.382(3) |
| C(40)-H(40) | 0.9500 |
| C(41)-C(42) | 1.390(4) |
| C(41)-H(41) | 0.9500 |
| C(42)-C(43) | 1.383(4) |
| C(42)-H(42) | 0.9500 |
| C(43)-C(44) | 1.389(3) |
| C(43)-H(43) | 0.9500 |
| C(44)-H(44) | 0.9500 |
| C(45)-C(50) | 1.397(3) |
| C(45)-C(46) | 1.398(3) |
| C(46)-C(47) | 1.390(3) |
| C(46)-H(46) | 0.9500 |
| C(47)-C(48) | 1.385(4) |
| C(47)-H(47) | 0.9500 |
| C(48)-C(49) | 1.381(4) |
| C(48)-H(48) | 0.9500 |
| C(49)-C(50) | 1.387(3) |
| C(49)-H(49) | 0.9500 |
| C(50)-H(50) | 0.9500 |
| C(18)-C(1)-C(2) | 119.24(19) |
| C(18)-C(1)-C(19) | 124.44(18) |
| C(2)-C(1)-C(19) | 116.28(18) |
| C(3)-C(2)-C(1) | 122.92(19) |
| C(3)-C(2)-H(2) | 118.5 |
| C(1)-C(2)-H(2) | 118.5 |
| C(2)-C(3)-C(4) | 118.40(19) |
| C(2)-C(3)-C(39) | 116.84(18) |
| C(4)-C(3)-C(39) | 124.71(18) |
| C(17)-C(4)-C(3) | 118.31(18) |
| C(17)-C(4)-C(5) | 117.64(18) |
| C(3)-C(4)-C(5) | 123.97(19) |
| C(6)-C(5)-C(10) | 118.32(19) |
| C(6)-C(5)-C(4) | 121.07(19) |
| C(10)-C(5)-C(4) | 120.33(19) |
| C(7)-C(6)-C(5) | 120.8(2) |
| C(7)-C(6)-H(6) | 119.6 |
| C(5)-C(6)-H(6) | 119.6 |
| C(6)-C(7)-C(8) | 121.5(2) |
| C(6)-C(7)-Br(1) | 119.4(3) |
| C(8)-C(7)-Br(1) | 119.1(3) |
| C(6)-C(7)-I(1A) | 115.7(2) |
| C(8)-C(7)-I(1A) | 122.8(2) |
| C(9)-C(8)-C(7) | 118.3(2) |
| C(9)-C(8)-H(8) | 120.8 |
| C(7)-C(8)-H(8) | 120.8 |
| C(8)-C(9)-C(10) | 121.8(2) |
| C(8)-C(9)-H(9) | 119.1 |
| C(10)-C(9)-H(9) | 119.1 |
| C(9)-C(10)-C(5) | 119.1(2) |
| C(9)-C(10)-C(11) | 121.9(2) |
| C(5)-C(10)-C(11) | 118.98(19) |
| C(12)-C(11)-C(16) | 118.7(2) |
| C(12)-C(11)-C(10) | 122.0(2) |
| C(16)-C(11)-C(10) | 119.22(19) |
| C(13)-C(12)-C(11) | 121.9(2) |
| C(13)-C(12)-H(12) | 119.1 |
| C(11)-C(12)-H(12) | 119.1 |
| C(12)-C(13)-C(14) | 118.6(2) |
| C(12)-C(13)-H(13) | 120.7 |
| C(14)-C(13)-H(13) | 120.7 |
| C(15)-C(14)-C(13) | 121.5(2) |
| C(15)-C(14)-Br(1A) | 122.0(2) |
| C(13)-C(14)-Br(1A) | 116.5(2) |
| C(15)-C(14)-I(1) | 116.94(19) |
| C(13)-C(14)-I(1) | 121.54(19) |
| C(14)-C(15)-C(16) | 120.5(2) |
| C(14)-C(15)-H(15) | 119.7 |
| C(16)-C(15)-H(15) | 119.7 |
| C(15)-C(16)-C(11) | 118.62(19) |
| C(15)-C(16)-C(17) | 119.76(19) |
| C(11)-C(16)-C(17) | 120.66(19) |
| C(4)-C(17)-C(18) | 120.02(18) |
| C(4)-C(17)-C(16) | 117.46(18) |
| C(18)-C(17)-C(16) | 122.20(19) |
| C(1)-C(18)-C(17) | 118.60(19) |
| C(1)-C(18)-C(45) | 121.54(18) |
| C(17)-C(18)-C(45) | 119.09(18) |
| C(28)-C(19)-C(20) | 119.67(19) |
| C(28)-C(19)-C(1) | 121.65(18) |
| C(20)-C(19)-C(1) | 118.35(18) |
| C(21)-C(20)-C(19) | 121.2(2) |
| C(21)-C(20)-H(20) | 119.4 |
| C(19)-C(20)-H(20) | 119.4 |
| C(20)-C(21)-C(22) | 121.2(2) |
| C(20)-C(21)-H(21) | 119.4 |
| C(22)-C(21)-H(21) | 119.4 |
| C(21)-C(22)-C(23) | 122.3(2) |
| C(21)-C(22)-C(27) | 118.70(19) |
| C(23)-C(22)-C(27) | 119.0(2) |
| C(24)-C(23)-C(22) | 120.7(2) |
| C(24)-C(23)-H(23) | 119.6 |
| C(22)-C(23)-H(23) | 119.6 |
| C(23)-C(24)-C(25) | 120.3(2) |
| C(23)-C(24)-H(24) | 119.9 |
| C(25)-C(24)-H(24) | 119.9 |
| C(26)-C(25)-C(24) | 121.0(2) |
| C(26)-C(25)-H(25) | 119.5 |
| C(24)-C(25)-H(25) | 119.5 |
| C(25)-C(26)-C(27) | 120.4(2) |
| C(25)-C(26)-H(26) | 119.8 |
| C(27)-C(26)-H(26) | 119.8 |
| C(26)-C(27)-C(22) | 118.55(19) |
| C(26)-C(27)-C(28) | 121.8(2) |
| C(22)-C(27)-C(28) | 119.61(19) |
| C(19)-C(28)-C(27) | 119.57(19) |
| C(19)-C(28)-C(29) | 120.51(18) |
| C(27)-C(28)-C(29) | 119.83(18) |
| C(38)-C(29)-C(30) | 119.8(2) |
| C(38)-C(29)-C(28) | 118.79(19) |
| C(30)-C(29)-C(28) | 121.36(19) |
| C(31)-C(30)-C(29) | 122.8(2) |
| C(31)-C(30)-C(35) | 118.3(2) |
| C(29)-C(30)-C(35) | 118.9(2) |
| C(32)-C(31)-C(30) | 120.9(2) |
| C(32)-C(31)-H(31) | 119.6 |
| C(30)-C(31)-H(31) | 119.6 |
| C(31)-C(32)-C(33) | 120.7(2) |
| C(31)-C(32)-H(32) | 119.7 |
| C(33)-C(32)-H(32) | 119.7 |
| C(34)-C(33)-C(32) | 119.9(2) |
| C(34)-C(33)-H(33) | 120.0 |
| C(32)-C(33)-H(33) | 120.0 |
| C(33)-C(34)-C(35) | 121.1(2) |
| C(33)-C(34)-H(34) | 119.4 |
| C(35)-C(34)-H(34) | 119.4 |
| C(36)-C(35)-C(34) | 121.6(2) |
| C(36)-C(35)-C(30) | 119.3(2) |
| C(34)-C(35)-C(30) | 119.0(2) |
| C(37)-C(36)-C(35) | 120.7(2) |
| C(37)-C(36)-H(36) | 119.7 |
| C(35)-C(36)-H(36) | 119.7 |
| C(36)-C(37)-C(38) | 120.1(2) |
| C(36)-C(37)-H(37) | 119.9 |
| C(38)-C(37)-H(37) | 119.9 |
| C(29)-C(38)-C(37) | 121.1(2) |
| C(29)-C(38)-H(38) | 119.5 |
| C(37)-C(38)-H(38) | 119.5 |
| C(44)-C(39)-C(40) | 118.9(2) |
| C(44)-C(39)-C(3) | 119.16(19) |
| C(40)-C(39)-C(3) | 121.98(19) |
| C(41)-C(40)-C(39) | 120.8(2) |
| C(41)-C(40)-H(40) | 119.6 |
| C(39)-C(40)-H(40) | 119.6 |
| C(40)-C(41)-C(42) | 119.6(2) |
| C(40)-C(41)-H(41) | 120.2 |
| C(42)-C(41)-H(41) | 120.2 |
| C(43)-C(42)-C(41) | 120.4(2) |
| C(43)-C(42)-H(42) | 119.8 |
| C(41)-C(42)-H(42) | 119.8 |
| C(42)-C(43)-C(44) | 119.8(2) |
| C(42)-C(43)-H(43) | 120.1 |
| C(44)-C(43)-H(43) | 120.1 |
| C(43)-C(44)-C(39) | 120.6(2) |
| C(43)-C(44)-H(44) | 119.7 |
| C(39)-C(44)-H(44) | 119.7 |
| C(50)-C(45)-C(46) | 117.9(2) |
| C(50)-C(45)-C(18) | 117.30(19) |
| C(46)-C(45)-C(18) | 124.83(19) |
| C(47)-C(46)-C(45) | 120.9(2) |
| C(47)-C(46)-H(46) | 119.5 |
| C(45)-C(46)-H(46) | 119.5 |
| C(48)-C(47)-C(46) | 120.1(2) |
| C(48)-C(47)-H(47) | 120.0 |
| C(46)-C(47)-H(47) | 120.0 |
| C(49)-C(48)-C(47) | 119.9(2) |
| C(49)-C(48)-H(48) | 120.1 |
| C(47)-C(48)-H(48) | 120.1 |
| C(48)-C(49)-C(50) | 120.1(2) |
| C(48)-C(49)-H(49) | 120.0 |
| C(50)-C(49)-H(49) | 120.0 |
| C(49)-C(50)-C(45) | 121.2(2) |
| C(49)-C(50)-H(50) | 119.4 |

________________________________________________________________

**Supplementary Table S4 |** Anisotropic displacement parameters (Å$^2$ x 10$^3$) for **2**. The anisotropic displacement factor exponent takes the form: $-2p^2[h^2a^{*2}U^{11} + ... + 2hk\,a^*b^*U^{12}]$.

___________________________________________________________________

|       | U$^{11}$ | U$^{22}$ | U$^{33}$ | U$^{23}$ | U$^{13}$ | U$^{12}$ |
|-------|-------|-------|-------|-------|-------|-------|
___________________________________________________________________

| | U$^{11}$ | U$^{22}$ | U$^{33}$ | U$^{23}$ | U$^{13}$ | U$^{12}$ |
|-------|-------|-------|-------|-------|-------|-------|
| C(1)  | 15(1) | 21(1) | 11(1) | 9(1)  | 3(1)  | 3(1)  |
| C(2)  | 19(1) | 13(1) | 13(1) | 4(1)  | -1(1) | 1(1)  |
| C(3)  | 14(1) | 21(1) | 12(1) | 7(1)  | 2(1)  | 3(1)  |
| C(4)  | 15(1) | 18(1) | 14(1) | 8(1)  | 5(1)  | 4(1)  |
| C(5)  | 15(1) | 18(1) | 20(1) | 11(1) | 7(1)  | 3(1)  |
| C(6)  | 17(1) | 24(1) | 22(1) | 14(1) | 7(1)  | 5(1)  |
| C(7)  | 17(1) | 32(1) | 28(1) | 20(1) | 6(1)  | 5(1)  |
| C(8)  | 20(1) | 25(1) | 41(1) | 21(1) | 11(1) | 9(1)  |
| C(9)  | 23(1) | 18(1) | 33(1) | 10(1) | 11(1) | 4(1)  |
| C(10) | 17(1) | 19(1) | 27(1) | 12(1) | 11(1) | 3(1)  |
| C(11) | 17(1) | 17(1) | 22(1) | 7(1)  | 8(1)  | -1(1) |
| C(12) | 22(1) | 16(1) | 32(1) | 5(1)  | 10(1) | 2(1)  |
| C(13) | 28(1) | 18(1) | 26(1) | -1(1) | 9(1)  | -5(1) |
| C(14) | 24(1) | 22(1) | 17(1) | 5(1)  | 2(1)  | -6(1) |
| C(15) | 20(1) | 18(1) | 17(1) | 6(1)  | 5(1)  | -2(1) |
| C(16) | 16(1) | 18(1) | 18(1) | 7(1)  | 6(1)  | 0(1)  |
| C(17) | 17(1) | 17(1) | 14(1) | 8(1)  | 3(1)  | 1(1)  |
| C(18) | 16(1) | 20(1) | 12(1) | 7(1)  | 2(1)  | 3(1)  |
| C(19) | 12(1) | 16(1) | 18(1) | 7(1)  | -1(1) | -1(1) |
| C(20) | 16(1) | 24(1) | 18(1) | 10(1) | 2(1)  | -1(1) |
| C(21) | 16(1) | 28(1) | 29(1) | 21(1) | -3(1) | -2(1) |
| C(22) | 12(1) | 17(1) | 33(1) | 14(1) | -2(1) | -2(1) |
| C(23) | 22(1) | 23(1) | 45(2) | 19(1) | -4(1) | 4(1)  |
| C(24) | 22(1) | 20(1) | 49(2) | 9(1)  | -1(1) | 7(1)  |
| C(25) | 20(1) | 20(1) | 38(1) | 2(1)  | 2(1)  | 5(1)  |
| C(26) | 15(1) | 18(1) | 26(1) | 4(1)  | 0(1)  | 1(1)  |
| C(27) | 11(1) | 15(1) | 25(1) | 7(1)  | 0(1)  | -1(1) |
| C(28) | 12(1) | 16(1) | 19(1) | 7(1) | 1(1) | 0(1) |
| C(29) | 14(1) | 15(1) | 17(1) | 4(1) | 6(1) | 6(1) |
| C(30) | 17(1) | 17(1) | 16(1) | 4(1) | 6(1) | 6(1) |
| C(31) | 19(1) | 20(1) | 23(1) | 7(1) | 5(1) | 4(1) |
| C(32) | 22(1) | 22(1) | 27(1) | 2(1) | 5(1) | 2(1) |
| C(33) | 23(1) | 32(1) | 20(1) | -3(1) | -3(1) | 5(1) |
| C(34) | 26(1) | 32(1) | 17(1) | 4(1) | 2(1) | 13(1) |
| C(35) | 23(1) | 22(1) | 16(1) | 7(1) | 7(1) | 11(1) |
| C(36) | 26(1) | 28(1) | 19(1) | 12(1) | 7(1) | 11(1) |
| C(37) | 25(1) | 22(1) | 25(1) | 13(1) | 8(1) | 5(1) |
| C(38) | 17(1) | 17(1) | 19(1) | 7(1) | 5(1) | 3(1) |
| C(39) | 17(1) | 16(1) | 21(1) | 11(1) | -1(1) | 3(1) |
| C(40) | 20(1) | 24(1) | 21(1) | 14(1) | 2(1) | 3(1) |
| C(41) | 16(1) | 35(1) | 35(1) | 22(1) | 3(1) | 1(1) |
| C(42) | 20(1) | 32(1) | 39(1) | 15(1) | -7(1) | -8(1) |
| C(43) | 25(1) | 20(1) | 27(1) | 6(1) | -6(1) | -2(1) |
| C(44) | 19(1) | 18(1) | 23(1) | 8(1) | 1(1) | 4(1) |
| C(45) | 17(1) | 17(1) | 13(1) | 3(1) | 2(1) | 3(1) |
| C(46) | 23(1) | 25(1) | 23(1) | 12(1) | -2(1) | 0(1) |
| C(47) | 20(1) | 35(1) | 29(1) | 14(1) | -5(1) | 2(1) |
| C(48) | 17(1) | 41(2) | 31(1) | 12(1) | -2(1) | -5(1) |
| C(49) | 22(1) | 35(1) | 32(1) | 18(1) | 3(1) | -6(1) |
| C(50) | 20(1) | 28(1) | 20(1) | 11(1) | 0(1) | 1(1) |
| Br(1) | 32(1) | 44(1) | 33(1) | 26(1) | 7(1) | 19(1) |
| I(1) | 42(1) | 25(1) | 16(1) | 8(1) | -8(1) | -4(1) |
| Br(1A) | 42(1) | 25(1) | 16(1) | 8(1) | -8(1) | -4(1) |
| I(1A) | 32(1) | 44(1) | 33(1) | 26(1) | 7(1) | 19(1) |

______________________________________________________________________

**Supplementary Table S5 |** Anisotropic displacement parameters (Å$^2$ x 10$^3$) for **2**. The anisotropic displacement factor exponent takes the form: $-2p^2 [ h^2 a^{*2} U^{11} + ... + 2 h k a^* b^* U^{12} ]$.

______________________________________________________________________

|       | x     | y     | z     | U(eq) |
|-------|-------|-------|-------|-------|

______________________________________________________________________

| H(2)  | 5612  | 6573  | 7488  | 19 |
| H(6)  | 7673  | 3803  | 7859  | 23 |
| H(8)  | 8615  | 446   | 6487  | 31 |
| H(9)  | 7155  | 26    | 5148  | 30 |
| H(12) | 6090  | -198  | 3824  | 30 |
| H(13) | 4659  | -544  | 2485  | 32 |
| H(15) | 3406  | 2609  | 4189  | 23 |
| H(20) | 3294  | 6216  | 5460  | 23 |
| H(21) | 2096  | 7721  | 5759  | 26 |
| H(23) | 986   | 9341  | 6995  | 35 |
| H(24) | 440   | 10417 | 8686  | 38 |
| H(25) | 1117  | 9990  | 10089 | 34 |
| H(26) | 2277  | 8441  | 9804  | 26 |
| H(31) | 4793  | 8469  | 9144  | 25 |
| H(32) | 6341  | 9352  | 10381 | 31 |
| H(33) | 6849  | 8582  | 11620 | 35 |
| H(34) | 5724  | 7009  | 11680 | 32 |
| H(36) | 4048  | 5477  | 10967 | 27 |
| H(37) | 2513  | 4559  | 9726  | 27 |
| H(38) | 2105  | 5205  | 8379  | 21 |
| H(40) | 8419  | 4660  | 6278  | 24 |
| H(41) | 10257 | 5524  | 7089  | 31 |
| H(42) | 10404 | 6892  | 8852  | 37 |
| H(43) | 8714  | 7450  | 9777  | 31 |
| H(44) | 6871  | 6576  | 8966  | 24 |
| H(46) | 2117  | 4481  | 4966  | 28 |
| H(47) | 203   | 3591  | 4617  | 33 |
| H(48) | -387  | 2264  | 5345  | 37 |
| H(49) | 948 | 1808 | 6402 | 35 |
| H(50) | 2866 | 2675 | 6737 | 27 |

____________________________________________________________________